\documentclass{PoS}
\usepackage{cite}

\graphicspath{{Figs/}}

\def\eq#1{Eq.~(\ref{#1})}

\def\fig#1{Fig.~\ref{#1}}

\def\sect#1{Sect.~\ref{#1}}
\def\vev{{\it vev}\ }

\title{Beyond the Standard Model: Charting Fundamental Interactions via Lattice Simulations}

\ShortTitle{Lattice Beyond the Standard Model}

\author{\speaker{Claudio Pica}\\
        CP$^3$-Origins, University of Southern Denmark,\\ Campusvej 55, DK-5230 Odense M, Denmark\\
        E-mail: \email{pica@cp3.sdu.dk}}

\abstract{After the discovery of the Higgs boson, the primary objective of the Large Hadron Collider (LHC) experiments is to identify new physics beyond the Standard Model (SM). One of the most intriguing possibilities would be the discovery of non-perturbative phenomena in electroweak physics.
In fact both ATLAS and CMS are providing crucial precision tests of the Higgs sector.
Most strikingly, there is no conclusive evidence yet on whether the Higgs boson is elementary or composite.

Lattice simulations can play a key role in advancing our theoretical understanding of strongly coupled gauge theories relevant for extensions of the SM and the LHC program. In this talk I will review the state of beyond the SM (BSM) lattice studies aimed to chart the phase diagram and to uncover the properties of strongly coupled gauge theories.
\\[.5cm]
\footnotesize{\it Preprint: CP3-Origins-2017-003 DNRF90}
}

\FullConference{34th annual International Symposium on Lattice Field Theory\\
		24-30 July 2016\\
		University of Southampton, UK}

\begin{document}

\section{Introduction}
\begin{figure}[b!]
     \center\includegraphics[width=.5\textwidth]{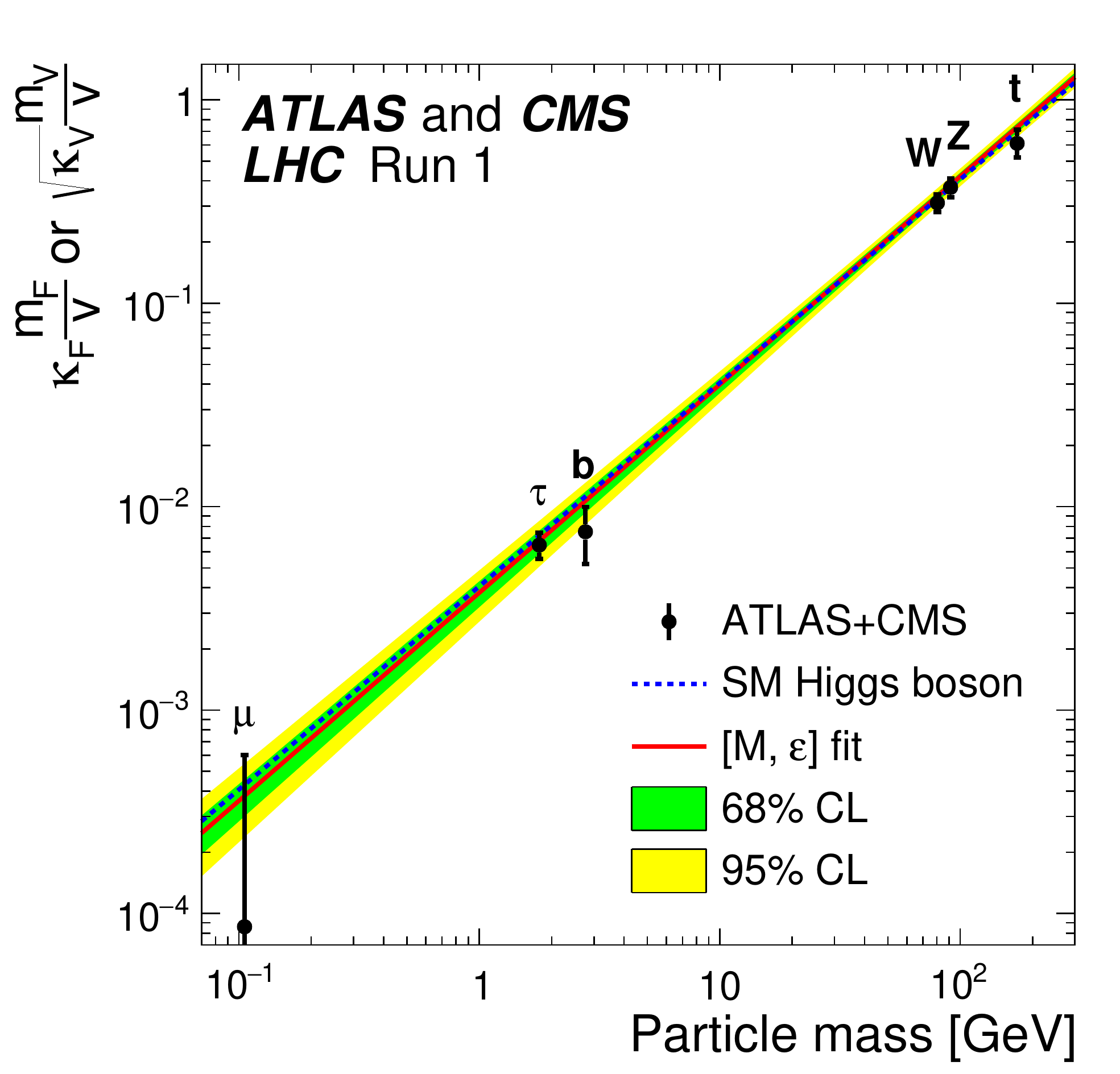}
     \caption{Best fit values as a function of particle mass for the combination of ATLAS and CMS data in the case of the couplings to the Higgs parametrized as $\kappa_F m_F/v$ for SM fermions and $\kappa_V m_V/v$ for weak vector bosons, where $v=246\mathrm{GeV}$. The dashed line is the prediction from the SM Higgs boson. From Ref.\cite{Khachatryan:2016vau}.}
     \label{fig:higgscouplings}
     \end{figure}

In 2012 the ATLAS and CMS experiments announced the discovery of a new particle, the Higgs boson, almost fifty years after its existence was first postulated. Since then, the LHC experiments are providing precise measurements of the properties of the newly discovered particle with the aim to establish its true nature, while at the same time searching for hints of new beyond the SM physics.
While no signs of new physics have yet been found, the properties of the Higgs particle are being explored in great detail. The favored quantum numbers for the new particle are the ones predicted by the SM, i.e. a CP-even, spin-$0$ scalar state. The mass of the Higgs boson is measured to be: $m_H=125.09\pm 0.21\mathrm{ (stat)}\pm0.11\mathrm{ (syst) GeV}$. The couplings of the new state to other SM fermions and vector bosons are also being measured precisely~\cite{Khachatryan:2016vau}. Several tests of the consistency of the Higgs couplings with the SM can be performed, based on different assumptions, showing no significant deviations from the predictions of the SM. \fig{fig:higgscouplings} from \cite{Khachatryan:2016vau} shows an example of such analysis which assumes all couplings to fermions and weak vector bosons rescaled by a universal factors $\kappa_F$ and $\kappa_V$ respectively.

Given the current experimental evidence of a light scalar particle with couplings compatible with the SM, it is tempting to conclude that its last missing piece has been found and the SM is valid  up to the Plank scale.
While attractive, there are several theoretical and experimental reasons which indicate the SM is incomplete.
On the experimental side, the SM does not account for the neutrino masses and the existence of dark matter nor does it provide an explanation for the origin of the matter-antimatter asymmetry of the Universe.
In addition some tensions exist between SM predictions and a few experimental measures such as the anomalous magnetic moment of the muon or the proton radius.
From a more theoretical perspective, the Higgs sector of SM is by many regarded as a convenient parametrization rather than a fundamental theory of electroweak symmetry breaking. Moreover, the elementary scalar nature of the Higgs field comes with an intrinsic instability of its mass under radiative corrections, giving rise to the hierarchy problem.

The two most solid proposals for resolving the Higgs naturalness problem are supersymmetry and compositeness. Supersymmetric models allow to maintain the perturbativity of the SM, at the expense of doubling the field content of the SM. Supersymmetric models are the main target for searches of new physics by the LHC experiments, but no evidence has been found for the existence of supersymmetric particles yet.
The second possibility is that the Higgs boson is a composite bound state of a new strong force. This solves the naturalness problem of the Higgs mass in the same way as the mass of the pion is natural in QCD: the radiative corrections to the pion mass are screened at scales of the order of $\Lambda_{QCD}$ as, at those scales, the pion ``dissolves'' into its UV degrees of freedom, quarks and gluons.
The composite nature of the Higgs is a more economical solution in terms of new fields, but it requires the presence of a new strongly-coupled sector in the model.

Here I will focus on this second possibility and, in particular, models of pseudo-Nambu-Goldstone boson (pNGB) Higgs or walking Technicolor (WTC) models.
Lattice studies provide a systematic and non-perturbative method to understand the new strong dynamics needed for composite Higgs models and can therefore provide valuable input for model building and the experimental searches at the LHC.
I will summarize the theoretical framework for composite Higgs models in \sect{sect:composite}, and review the recent efforts of the lattice community presented at this conference in \sect{sect:cwindow} (conformal window and the search for infrared conformal models with large anomalous dimensions),~\sect{sect:scalars} (searches for walking Technicolor models with light scalars) and \sect{sect:pngbhiggs} (non-perturbative studies of pNBG models).

Although composite Higgs models are the primary focus of the lattice BSM community, many other topics have also been presented which I will not discuss here, such as supersymmetry \cite{Kamata:2016,giedt:2016,schaich:2016,august:2016,joseph:2016,giudice:2016}, extradimensions \cite{alberti:2016}, gauge/gravity duality \cite{bennett:2016,forini:2016,berkowitz:2016} and asymptotically safe gauge-Yukawa theories \cite{buy:2016,bond:2016}. For a review of recent results for dark matter models from lattice simulations see \cite{rinaldi:2016}.

\section{Composite Higgs models}\label{sect:composite}
Within the SM, electroweak symmetry breaking (EWSB) is assumed but it has no dynamical origin.
The composite Higgs idea is an interesting realization of EWSB via a new strong dynamics.

\begin{figure}[t!]
     \center\includegraphics[width=.6\textwidth]{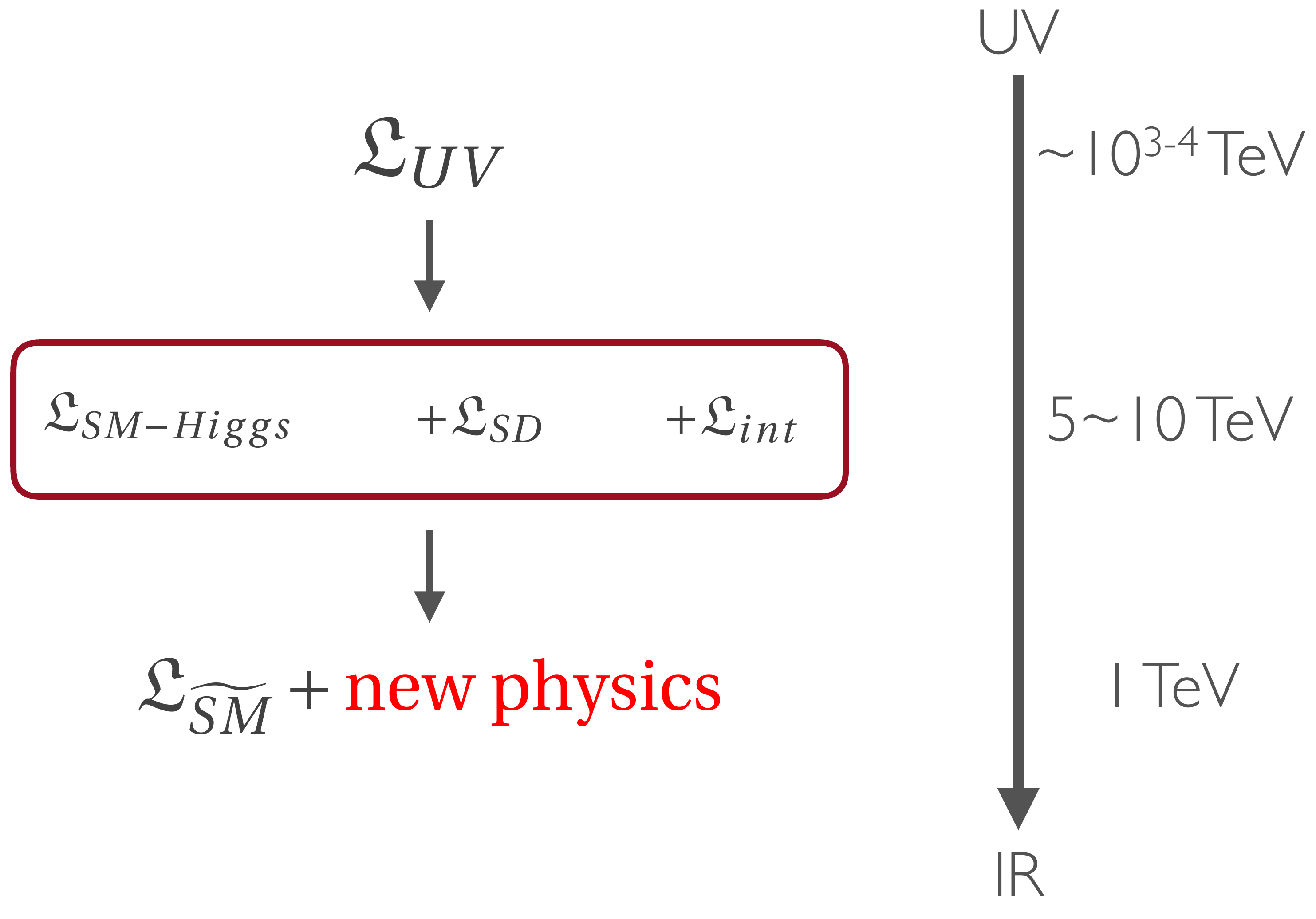}
     \caption{Structure of a composite Higgs model as described in the main text.}
     \label{fig:chmodel}
     \end{figure}

Composite Higgs models are built starting from the SM by removing the elementary Higgs field and replacing it with a new strongly interacting sector. This new strong sector is required to break electroweak symmetry dynamically, to give the correct mass to the W and Z bosons and to feature a composite scalar Higgs particle that mimics the SM one. If we require the new strong sector to be UV complete, the simplest realizations are based on non-abelian gauge theories with only fermionic matter. This new strong sector will possess a full spectrum of resonances, some of which could also be relevant in the context of dark matter models.
The scale of the new strong interaction is set at the electroweak scale by the requirement that the W and Z bosons should acquire the experimentally observed mass.

In order to generate masses for the SM fermions, additional interactions are required in any realistic model. These new interactions can be included in the model as effective operators of the form of four-fermion operators, stemming from some unspecified UV dynamics at a much high energy scale.
A cartoon of the structure of a generic composite Higgs model is sketched in \fig{fig:chmodel}, where ${\cal L}_{SM-Higgs}$ is the Lagrangian describing the SM without the Higgs field, ${\cal L}_{SD}$ the new strong sector, and ${\cal L}_{int}$ is the effective Lagrangian describing the interactions required to generate SM fermion masses.
At the electroweak scale, the composite Higgs model is constrained by experimental data to closely resemble the SM, but it will feature additional new resonance states, which are composites from the new strong sector.
To avoid conflict with electroweak precision data, it is necessary that a mass gap separates the Higgs resonance from the other resonances of the strong sector.

It is worth to stress that in lattice simulations only the new strong sector in isolation is studied, while in any realistic model the SM and other additional interactions will affect the dynamics of the new strong sector.
In particular the composite Higgs properties are expected to be affected significantly by these interactions (see below), while heavier resonances should be less affected.
When comparing lattice simulations to experiments, it is therefore important to consider only observables which only depend on the strong sector or to estimate the corrections from the other interactions.

The two most interesting limits of composite Higgs models are Walking Technicolor and pNBG Higgs models. In such models the little hierarchy between the Higgs mass and the other resonances of the strong model is explained by an extra approximate global symmetry and the Higgs boson being the associated pNBG. This extra symmetry is a global flavor symmetry in the case of pNBG Higgs models, or an approximate scale invariance symmetry in the case of WTC.
The main features of WTC and pNBG models are summarized below.

\subsection{Walking Technicolor}
A time-honoured idea for dynamical electroweak symmetry breaking is walking techinicolor \cite{Weinberg:1979bn,Susskind:1978ms,Dimopoulos:1979es,Eichten:1979ah,Farhi:1980xs,Holdom:1984sk,Yamawaki:1985zg,Bando:1986bg,Appelquist:1987fc,Caswell:1974gg,Banks:1981nn}.
This idea was recently revived by extending the original framework by considering an enlarged theory space in the choice of the gauge group, number of flavors, and fermion representation \cite{Sannino:2004qp,Hong:2004td} which triggered considerable new interest in the lattice community, as non-perturbative methods are required to make precise statements about the properties of these strongly coupled models.
In recent years, most of the effort of the lattice community in BSM physics has been devoted to the study of the so-called ``conformal window'', motivated by WTC models.
\begin{figure}[t!]
     \center\includegraphics[width=.8\textwidth]{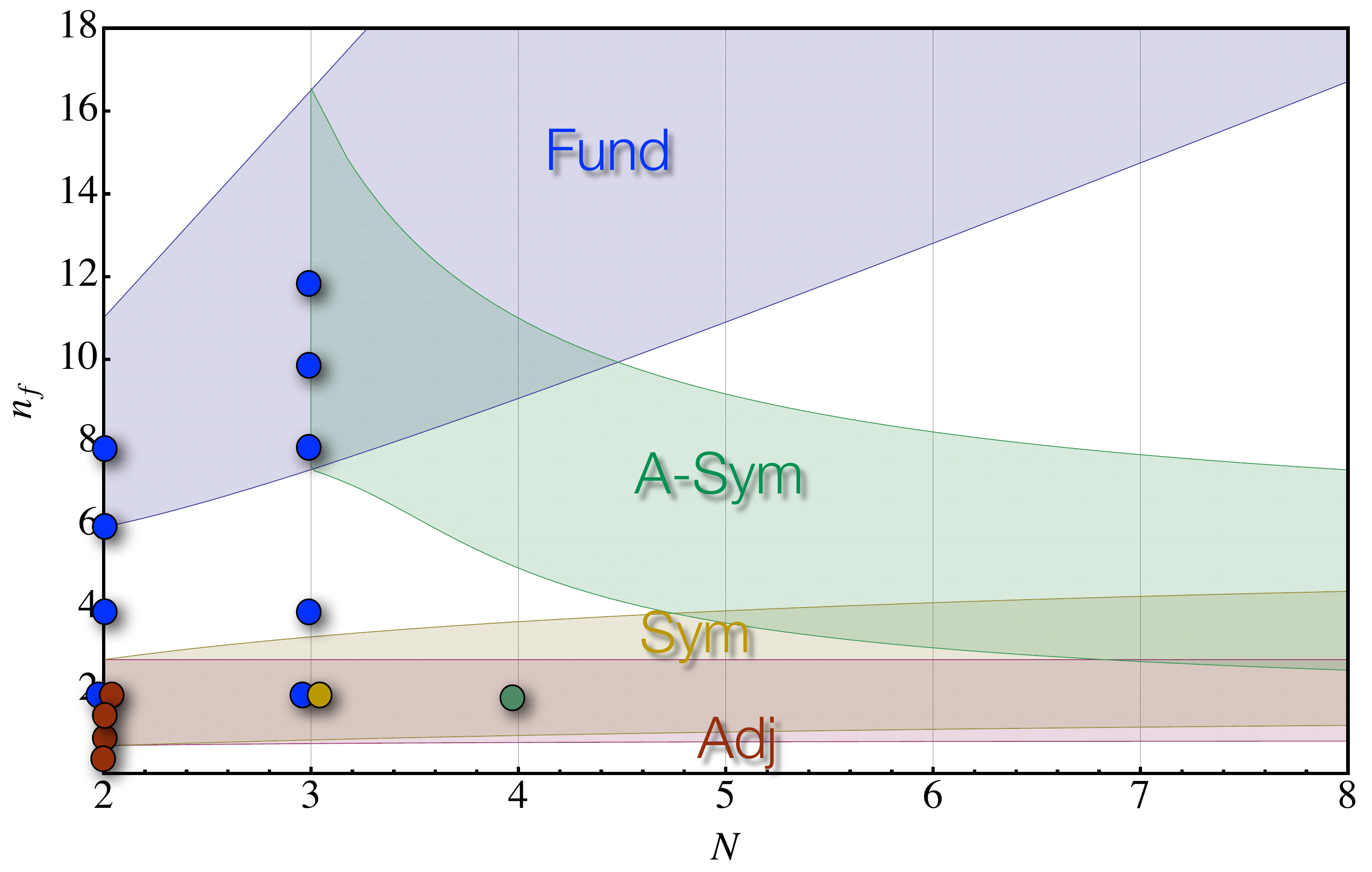}
     \caption{Chart of models of strong dynamics presented at Lattice 2016. Colors represent different representations of the SU(N) gauge group: fundamental, adjoint, two-index symmetric and antisymmetric. The continuous band is the perturbative 4-loop estimate ($\overline{\rm MS}$ scheme) of the conformal window from \cite{Pica:2010xq}. }
     \label{fig:chart}
     \end{figure}
In fact a number of four dimensional UV complete models are readily available to explore. I show in \fig{fig:chart} a chart of the ``theory space'' with the models of strong dynamics presented at the Lattice 2016 conference. The exploration of novel strong dynamics, different than QCD, is a major ongoing effort of the BSM lattice community, and many results are available for a number of interesting models.

Technicolor models feature a new strong sector typically based on a SU(N) gauge group, but Sp(2N), SO(N) and even exceptional simple Lie groups have also been considered~\cite{Sannino:2009aw,Mojaza:2012zd}, and $n_f$ massless techni-fermions in a given representation of the gauge group\footnote{WTC models with fermions in two different representations of the gauge group also exist~\cite{Ryttov:2008xe,Ryttov:2009yw} but they will not be addressed here.}.
It is required that the new strong dynamics features dynamical symmetry breaking leading to at least three Nambu-Goldstone boson and the formation of a techni-fermion condensate.
When electroweak interactions are introduced, the techni-fermion condensate breaks electroweak symmetry and the W and Z bosons acquire the correct mass, provided that the scale of the new strong force is chosen so that the techni-pion decay constant is equal to the electroweak \textit{vev}, $F_{\pi}\simeq 246$ GeV.
A minimal realization is obtained considering $n_f=2$ techni-fermions in a complex representation of the gauge group, such as the fundamental of SU(3) leading to a scaled-up version of QCD, whose pattern of spontaneous chiral symmetry breaking is SU(2)$\times$SU(2)/SU(2) which as the minimal number of NG bosons required for a TC model. For a larger number of techni-fermions, or for real or pseudoreal representations, TC models will in general lead to additional NG bosons in the physical spectrum.

While TC models provide a natural explanation of the EW scale, there are severe constraints for a realistic TC model. The TC composite Higgs particle is identified as the lightest scalar excitation of the condensate, which has to be a narrow, light resonance (unlike the equivalent $f_0(500)$ state in QCD). Moreover the couplings between the composite Higgs particle and the SM fermions must be SM-like.
Electroweak precision data, such as the Peskin-Takeuchi S and T parameters impose strict constraints on the model, which are not easy to satisfy.
Finally to generate SM fermion masses, additional interactions are needed in the form of extended TC interactions (ETC), which generically also generate flavor changing neutral current (FCNC) among SM particles which are experimentally very small.

The ``walking'' TC idea was introduced to alleviate the problems of TC.
Let's start by considering the problem of SM fermion mass generation. ETC interactions at some large UV scale $\Lambda_{ETC}$ will produce effective couplings among the techni-fermions and the SM fermions of three kinds:
\begin{eqnarray}
A_{ab}\frac{\overline{Q}T^a Q\overline{\psi}T^b \psi}{\Lambda_{ETC}^2}+B_{ab}\frac{\overline{Q}T^a Q\overline{Q}T^b Q}{\Lambda_{ETC}^2}+C_{ab}\frac{\overline{\psi}T^a \psi\overline{\psi}T^b \psi}{\Lambda_{ETC}^2}\ ,
\label{eq:etc}
\end{eqnarray}
where we have schematically indicated with $Q$ a generic techni-fermion, $\psi$ a SM fermion, $T$ the ETC gauge group generators and $A$, $B$, $C$ are adimensional coefficients of order assumed generically of order $1$.
The last term generates FCNC, and the experimental constraints from $B\bar B$ and $K\bar K$ mixings can be used to set a lower limit on $\Lambda_{ETC}$. Assuming no additional structure for the ETC interactions and coefficients $C\simeq 1$ one obtains a limit $\Lambda_{ETC}\gtrsim 10^3$ TeV for the second generation of SM quarks\cite{Bona:2007vi, Chivukula:2010tn, Andersen:2011yj}.
After techni-fermions have condensed, the $A$ terms in \eq{eq:etc} lead to mass terms for the SM fermions:
\begin{eqnarray}
    m_q\simeq \frac{1}{\Lambda_{ETC}^2}\langle\bar Q Q\rangle_{ETC}\, .\label{eq:qmass}
\end{eqnarray}
From this expression it is clear that in order to generate large hierarchies in the quark masses, the generation of the four-fermion operators for different SM flavors cannot just happen at one single ETC scale. Rather one should assume a different scale $\Lambda_{ETC}$ for each SM quark family.
From the FCNC constraint above, the charm and strange quark mass should be generated at a scale of $~10^3$ TeV, if no further suppression mechanism is present in the ETC model.
The techni-quark condensate in \eq{eq:qmass} is evaluated at the ETC scale and it is related to the condensate at the TC scale by the renormalization group equation: $\langle\bar Q Q\rangle_{ETC}=\langle\bar Q Q\rangle_{TC}\exp(\int_{\Lambda_{TC}}^{\Lambda_{ETC}}\gamma(\mu)d\ln(\mu))$, where $\gamma$ is the anomalous dimension of the techni-quark mass operator.
Assuming that the TC model is asymptotically free just above the TC $\sim$ 1 TeV, i.e. that $\gamma\sim 0$ above $\Lambda_{TC}$, and estimating by simple dimensional analysis $\langle\bar Q Q\rangle_{TC}\sim \Lambda_{TC}^3$, one obtains $m_q\sim\Lambda_{TC}(\Lambda_{TC}/\Lambda_{ETC})^2$. This results in a mass of $\sim 1$ MeV for the second generation of SM quarks, which is clearly too small.

To resolve this tension the mechanism of walking TC was proposed. Instead of assuming the TC dynamics to be QCD-like as done above, one requires the model have an approximate conformal symmetry, so that the anomalous dimension $\gamma(\mu)$ remains almost constant between the TC and ETC scales and equal to $\gamma^*$. Under this assumption the SM fermion mass can be estimated as $m_q\sim\Lambda_{TC}(\Lambda_{TC}/\Lambda_{ETC})^{2-\gamma^*}$. For large values of $\gamma^*\sim1$ this results in a considerable enhancement of the generated SM fermion mass $\sim 1$ GeV. A large $\gamma^*\sim1$ requires a strongly coupled (near) conformal dynamics. Whether or not this can be realized is a very interesting question which can in principle be answered by lattice simulations. I will discuss below in \sect{sect:cwindow} the current status for lattice searches of IR conformal models with large mass anomalous dimensions.

The $B$ terms in \eq{eq:etc}, four-fermion interactions among techni-quarks, will affect the strong TC dynamics\footnote{One could also use an infrared conformal TC model for the new strong sector, in which case four-fermions interactions could change the anomalous dimensions and also generate walking \cite{Fukano:2010yv,Rantaharju:2016zjj}}, and they will e.g. generate masses for the techni-pions.

Besides SM fermion mass generation, the Peskin-Takeuchi S-parameter also imposes a strict constraint on the TC dynamics. It has been suggested that the value of the S-parameter, normalized to the number of new electroweak doublets, is reduced in near-conformal models of walking TC~\cite{Appelquist:1998xf,Sannino:2010ca,DiChiara:2010xb}. Evidence from non-perturbative lattice simulations is still rather limited, see \cite{Appelquist:2014zsa} for a recent result for the SU(3) gauge group with $N_f=2,\,6,\,8$ flavors. This preliminary evidence favors the expected reduction of the S-parameter for walking TC models, although a precise determination is hindered by difficulties related to the chiral extrapolation in near-conformal models.

Walking TC models are also challenged by the requirement of a light composite Higgs scalar, with the correct couplings to the SM fermions. In TC models the composite Higgs is the lightest isospin-0 scalar composite of techni-quarks. In a strongly coupled model this composite state will also contain a techni-glue component. The observation that the QCD analogue, the $f_0(500)$ resonance, is a quite broad resonance, has driven the belief that light composite states cannot exist in a TC model. However this naive expectation does not take into account that any realistic WTC model should: 1) to take into account interactions with the SM particles; and 2) feature a near conformal dynamics quite different than QCD.

Interactions with SM particles will change the mass of the composite Higgs state. For a realistic model, assuming SM-like couplings to gauge bosons and SM fermions, the corrections are large and dominated by the negative top quark loop contribution\cite{Foadi:2012bb}, so that a mass as large as $\sim$1 TeV for the TC Higgs in isolation is not excluded. If the mass is reduced significantly by these interactions, the composite Higgs state would become narrow due to kinematics.
The assumption of SM-like coupling is a delicate issue and it is unclear that it should hold for a generic WTC model. In \cite{Foadi:2012bb} it was argued that couplings of the TC Higgs to the W and Z bosons are SM-like by comparison with the $\sigma\pi\pi$ effective coupling in QCD. A more detailed analysis in~\cite{Belyaev:2013ida} has confirmed this expectation and found the effective coupling in QCD to be in surprisingly good agreement with the corresponding SM coupling.
Couplings to SM fermions are more model dependent. Nonetheless if the SM fermion mass is generated via ETC interactions, the same four-fermion interactions will generate effective Yukawa couplings for the composite TC Higgs, which are then expected to be proportional to the SM fermion masses~\cite{Foadi:2012bb}.

Another interesting possibility has been considered which leads to a light composite Higgs with correct couplings to the SM particles, namely the possibility of a dilaton-Higgs~\cite{Yamawaki:1985zg,Dietrich:2005jn,Goldberger:2008zz,Bellazzini:2012vz}.
A dilaton-Higgs would be light, as the pNGB associated to an approximate scale invariance of the model, and with SM-like couplings, as the dilatonic states couples to the trace of the energy-momentum tensor, if the scale at which scale invariance is broken is the electroweak scale. Such a dilaton-Higgs would be difficult to distinguish from the SM Higgs~\cite{Bellazzini:2012vz}. If a near-conformal WTC model can produce such a light dilatonic scalar particle is still an open question, which lattice simulations are trying to address.

In recent years the lattice community has provided evidence for the existence of light composite scalar states in strongly coupled gauge theory, pointing against the common belief that such states cannot exist. I will summarize below in \sect{sect:scalars} the status of lattice searches for light composite scalars.

\subsection{pNGB Composite Higgs}
\label{sect:ngh}

pNGB composite Higgs models interpolate between TC models and the SM with a fundamental Higgs. As in TC, pNGB composite Higgs models start by assuming a new strongly interacting sector with a global symmetry group $G_F$ that is spontaneously broken to a subgroup $H_F$.

Unlike in TC, the Higgs particle is identified with a Nambu-Goldstone boson\cite{Kaplan:1983fs,Kaplan:1983sm,Banks:1984gj,Georgi:1984ef,Georgi:1984af,Dugan:1984hq} of the new strongly interacting sector. This naturally explains why the composite Higgs is light, and, since based on a similar mechanism, one can easily recover the correct SM-like coupling between the composite Higgs and the electroweak gauge bosons.

For a viable realization of this scenario, as long as only electroweak symmetry is concerned, the pattern of  symmetry breaking $G_F/H_F$ should be such that the custodial symmetry of the SM is preserved, i.e. $H_F\supset G_{\rm cust}$=SU(2)$_L\times$SU(2)$_R$, and that one of the Nambu-Goldstone bosons has the correct quantum numbers for the Higgs particle, i.e. belong to the irrep $(2,2)$ of $G_{\rm cust}$.

If we consider UV-complete models in four dimensions featuring fermionic matter, the three minimal cosets are SU(4)$\times$SU(4)/SU(4) for fermions in a complex representation of the gauge group, SU(4)/Sp(4) for fermions in a pseudoreal representation and SU(5)/SO(5) for fermions in a real representation. Models in all three of the minimal cosets have been considered at this Lattice conference.
Models based on the pattern SU(4)/Sp(4) contain only five NGB, i.e. the three required to generate masses for the W and Z bosons, one composite Higgs scalar, and one additional NGB.
This pattern of symmetry breaking is realized by a SU(2) technicolor gauge group with $N_f=2$ Dirac fermions in the fundamental representation, which therefore is the minimal realization of a UV-complete pNGB composite Higgs model\footnote{The SU(4)/Sp(4) is equivalent to the SO(6)/SO(5) coset, sometimes called next to minimal coset, which has been studied in detail via an effective sigma model description. The so-called minimal coset SO(5)/SO(4) lacks a four dimensional UV completion.}.
The familiar-looking SU(4)$\times$SU(4)/SU(4) coset can be realized for an SU(3) gauge group and four Dirac fermions in the fundamental representation, while the SU(5)/SO(5) coset can emerge from an SU(4) gauge group with five Majorana fermions in the two-index antisymmetric representation.

Electroweak interactions will break the global symmetry of the strong sector and generate a mass for the composite Higgs, in a similar way as electromagnetic interactions generate a mass for charged pions in massless QCD. This potential however will not generate a \vev for the composite Higgs field so that the minimum of the effective potential occurs at $\langle h\rangle=0$ and electroweak symmetry remains unbroken. To break electroweak symmetry interactions with SM fermions are needed, the dominant contribution coming from the top quark. Possibly other sources of explicit symmetry breaking for the global flavor symmetry of the new strong sector can be considered.

Interaction with the SM top quark will change the composite Higgs mass and generate a non-zero $\langle h\rangle$ and break electroweak symmetry.
The breaking can be parametrized by an angle $\theta$ so that: $\frac{v}{F_\pi}=\sin\theta=\sin \frac{\langle h\rangle}{F_\pi}$, where $v=$246 GeV and $F_\pi$ is the NGB decay constant of the strong sector.
Other composite strong resonances are expected at a scale $\sim 4\pi F_\pi$ in a QCD-like dynamics.
Therefore, while interactions with gauge bosons W and Z tend to align the vacuum angle $\theta$ towards zero, i.e. the NGB composite Higgs direction, SM fermions interactions have the opposite effect of pushing  towards large alignment angles in the Technicolor direction $\theta\simeq \pi/2$.
In the limit $\theta\to 0$,  the model naturally has couplings to the electroweak gauge boson identical to the ones of the SM: $g_{VVh}=g_{VVh}^{SM}\cos\theta$ and $g_{VVhh}=g_{VVhh}^{SM}\cos 2\theta$.
Realistic models require a light Higgs mass and SM-like couplings with the EW bosons, which implies rather small $\theta$  angles.
There is therefore a degree of fine tuning required between EW gauge bosons and SM fermions contributions, which are completely different in nature.

Coupling to SM fermions can be introduced either as in ETC, i.e. quadratic in the SM fermions $\frac{{\mathcal O}_S\overline{\psi} \psi}{\Lambda_{UV}^{d-1}}$, or in the partial compositeness way, i.e. linearly $\frac{{\mathcal O}_F \psi}{\Lambda_{UV}^{d-5/2}}$, where $d$ is the dimension of the composite scalar or fermionic operator ${\mathcal O}$.
As before, these interactions will generically also produce FCNC. To evade experimental constraints one then consider models where $\Lambda_{UV}$ can be pushed to a high scale, while still generating the required mass.
In the extreme case of vanishing exponent for $\Lambda_{UV}$, one can consider the case of a fermion bilinear for ${\mathcal O}_S$ as in ETC which implies an anomalous dimension of $\gamma=2$ and a three-quark "baryon" fermionic operator for ${\mathcal O}_F$ also implying $\gamma=2$ for its anomalous dimension\footnote{Both values are within the unitarity limit for conformal field theories, the limit for a scalar being 2 and for a fermion being 3.}.
Therefore, as in TC, large anomalous dimensions are naturally advocated in models of partial compositeness.
Whether such models of strong dynamics exists or not is still an open question.
Recently it was realized that such large anomalous dimensions for baryon operators in the simplest realization of partial compositeness, featuring only fermions in one representation of the gauge group, are very unlikely~\cite{Pica:2016rmv}.

Couplings of the composite Higgs to SM fermions are model dependent, however one can generically recover MS-like couplings in the limit $\theta\to 0$: $g_{h\bar{f}f}=g_{h\bar{f}f}^{SM}(1 + c \theta^2 + \ldots)$. The S-parameter are also model dependent but it can naively be estimated, via the so-called "zertoh" Weinberg sum rule, to be $S\propto (\frac{v}{m_\rho})^2\propto \sin^2\theta$, which shows that in the limit of small $\theta$ the contributions to the S-parameter are small, as the resonances from the strong sector become heavy.
The experimental constraint on the S-parameter poses one of the most stringent limit on the size of $\theta\lesssim 0.2-0.3$.

Models of partial compositeness require the composite fermions operators, which are the partners of SM quarks, to carry QCD color.
This in turn implies that the global symmetry of the new strong sector should be enlarged to allow the embedding of SU(3)$_c$.
Since models with partners for all the SM fermions are difficult to construct, a common approach is to use partial compositeness for the top sector only.
A classification of possible models featuring only fermionic matter in two different representation of the gauge group was given in \cite{Ferretti:2013kya,Ferretti:2016upr}, while a solution based on SU(3) hypercolor group and $n_f>6$ fermions in the fundamental representation can be found in \cite{Vecchi:2015fma}.
Among models with fermions in two different representations, an interesting one, obtained by extending the SU(5)/SO(5) coset, is based an SU(4) gauge group with five Majorana fermions in the two-index antisymmetric representation plus three Dirac fermions in the fundamental representation \cite{Ferretti:2016upr}.

Finally, recently models of partial compositeness based on a strong sector with both fermions and (hyper)colored scalars have been proposed~\cite{Sannino:2016sfx}, in which a mass for all SM fermions can be generated without requiring large anomalous dimensions. These can be seen as effective description for any partial compositeness model with large anomalous dimensions and in \cite{Sannino:2016sfx} it was shown that such models can be well-defined with small quartic scalar couplings up to the Plank scale.

\subsection{Non-perturbative questions for the lattice}
Lattice gauge theory simulations are a unique tools to understand strong dynamics, and they can be used to address a number of relevant questions about the non-perturbative dynamics of the strongly interacting composite Higgs models, such as:
\begin{itemize}
	\item Where is the exact location of the conformal window (see \fig{fig:chart})? i.e. which asymtotically free models have an IR fixed point?
	\item For models inside the conformal window, what is the (fixed point) anomalous dimension of the mass or the anomalous dimension of baryonic operators? Is there any model with large anomalous dimensions?
	\item For models outside the conformal window, can composite scalar states analogue to the $\sigma$ resonance be light and narrow in a strongly coupled dynamics? If so, is such state a dilaton? What are the couplings of such composite light scalar to NGBs?
	\item More generally, how does the spectrum of a strongly coupled model changes with the number of fermion field or when changing the number of colors or fermion representation? e.g. what is the ratio $m_\rho/F_\pi$ or  $m_\sigma/F_\pi$ ?
	\item How big is the S-parameter for models just below the conformal window?
\end{itemize}
It is the primary goal of the lattice BSM effort to answer such questions, and a number of interesting results for the models in \fig{fig:chart} is already available. I will summarize below the new lattice results presented at this conference.

\section{Conformal Window and anomalous dimensions}\label{sect:cwindow}

The precise location of the conformal window is a primary objective of lattice BSM studies.
The focus has been on models based on SU(2) and SU(3) gauge groups, with many fermions in the fundamental representation, but also an interesting update for models based on SU(2) with adjoint fermions has been presented at this conference.
I summarize below the main new results which show that models with large anomalous dimensions inside the CW have not yet been found despite the large number of models investigated so far.

\subsection{SU(3) with fundamental fermions\label{sect:su3f}}
The exact size of the conformal window in this case is still debated. There is a general consensus that $n_f=6$ lies outside the conformal window and most groups agree that $n_f=8$ is outside the CW\footnote{See however e.g. \cite{daSilva:2015vna} for an alternative point of view.} (studies on the spectrum of the $n_f=8$ model will be reported below in \sect{sect:scalars}).

\begin{figure}[t!]
     \center
     \includegraphics[width=.49\textwidth]{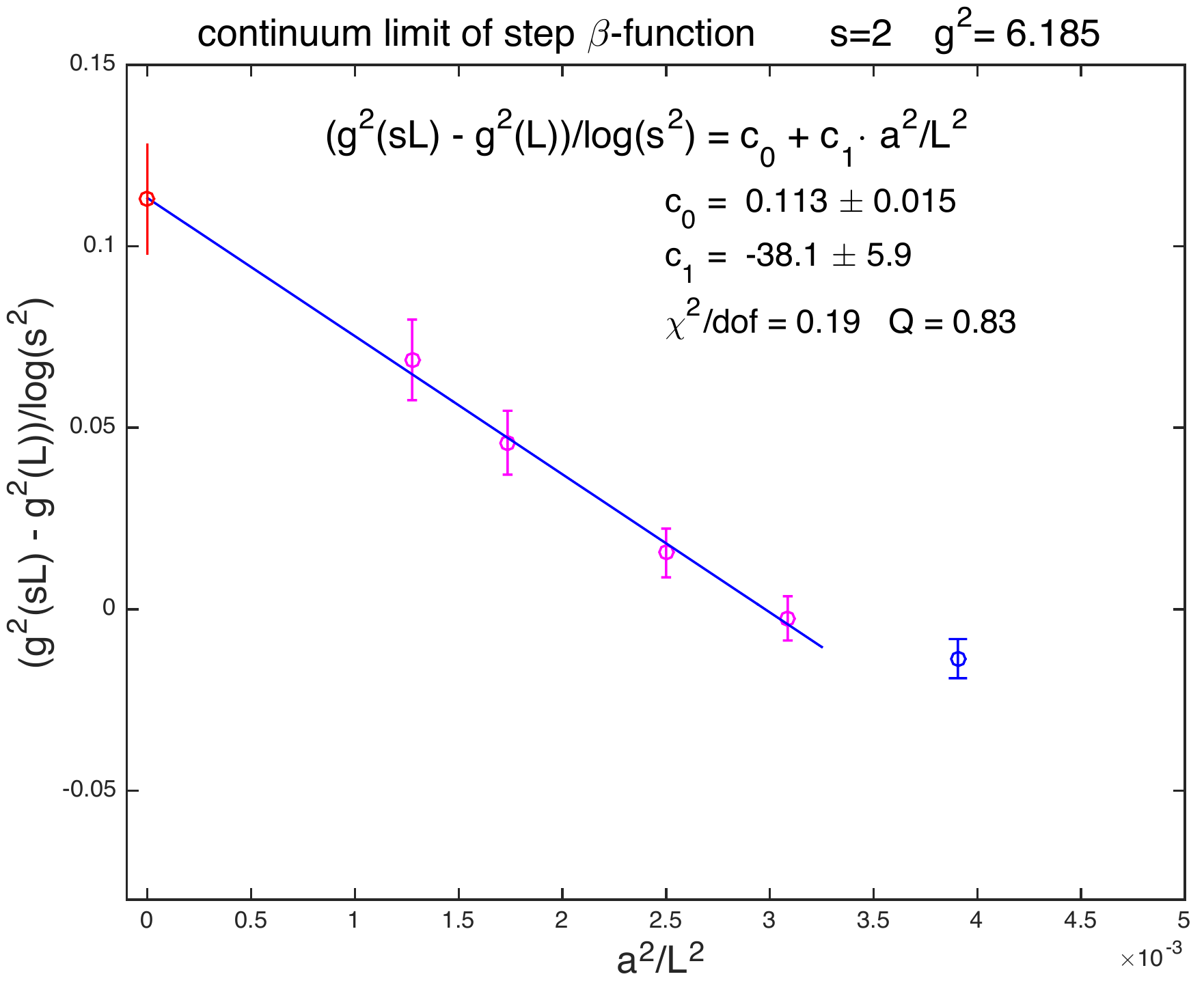}\hfill
     \includegraphics[width=.49\textwidth]{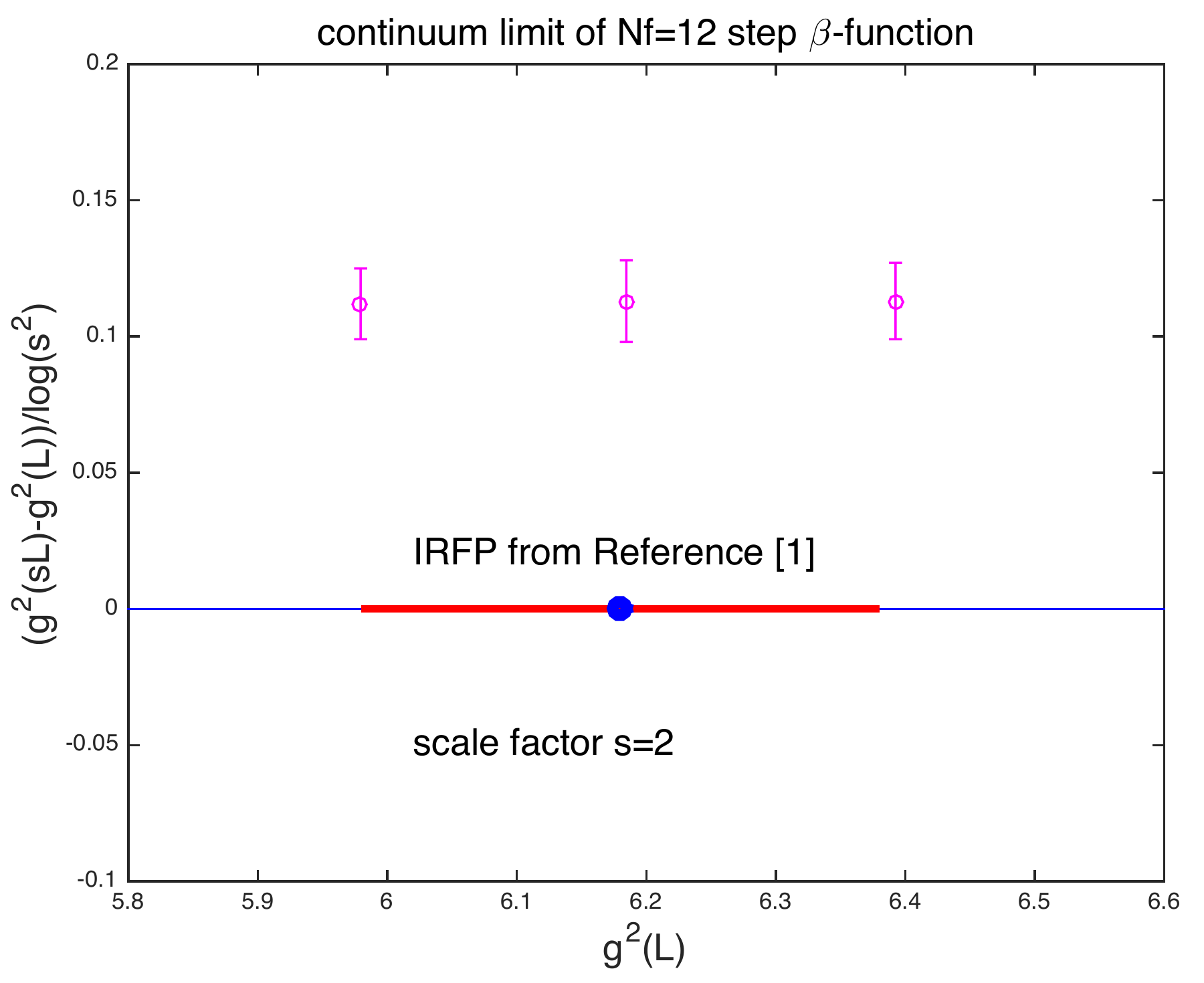}
     \caption{Left panel: example of continuum extrapolation of the discrete $\beta$-function using lattice volumes as large as $L/a=56$. Right panel: Continuum discrete $\beta$-function for the SU(3) $n_f=12$ model across the region where an IR fixed point, in the same scheme, was reported by previous state-of-the-art studies. From \cite{Fodor:2012td}. }
     \label{fig:nogradi}
     \end{figure}

New results have been presented for the cases of $10$ and $12$ fermions.
The $n_f=12$ case has been studied extensively by several groups by a variety of different methods~\cite{Appelquist:2009ty,Appelquist:2007hu, Lin:2012iw,Fodor:2011tu, Appelquist:2011dp, DeGrand:2011cu, Aoki:2012eq, Cheng:2013eu,Cheng:2013xha, Aoki:2013zsa, Lombardo:2014pda,Hasenfratz:2011xn,Cheng:2014jba}. While many of these studies indicate that the model is inside the conformal window and the anomalous dimension of the mass is small $\gamma^*\sim 0.2-0.3$, new evidence was presented in \cite{Fodor:2016zil}, and reported at this conference, against the existence of the IR fixed point at the location reported by previous lattice studies.
This new study is a high-precision measurement of the non-perturbative discrete $\beta$-function of the model in the gradient flow coupling and finite volume scheme~\cite{Fodor:2012td}.
The scheme was chosen to match, in the continuum, the one used in previous studies of the $n_f=12$ model so that the results can be easily compared. By using much larger lattice volumes up to $L/a=56$, and a precise tuning of the bare coupling, the quality of the continuum extrapolation was dramatically improved over previous state-of-the-art determinations.
In \fig{fig:nogradi} we show an example of continuum extrapolation  (left panel) and the final result of \cite{Fodor:2012td} (right panel) for the discrete $\beta$-function.
The new results are clearly incompatible with the existence of an IR fixed point where previously reported. The possibility that the $n_f=12$ model is IR conformal with a fixed point at a stronger coupling is also clearly not excluded.
This study proves the necessity of using very large lattice volumes in the vicinity of a candidate IR fixed point.
In the case of the determination of a non-perturbative $\beta$-function via a step-scaling procedure, the large lattice volume is then required to reach the correct continuum limit and avoid lattice artifacts.

\begin{figure}[t!]
     \center
     \includegraphics[width=.49\textwidth]{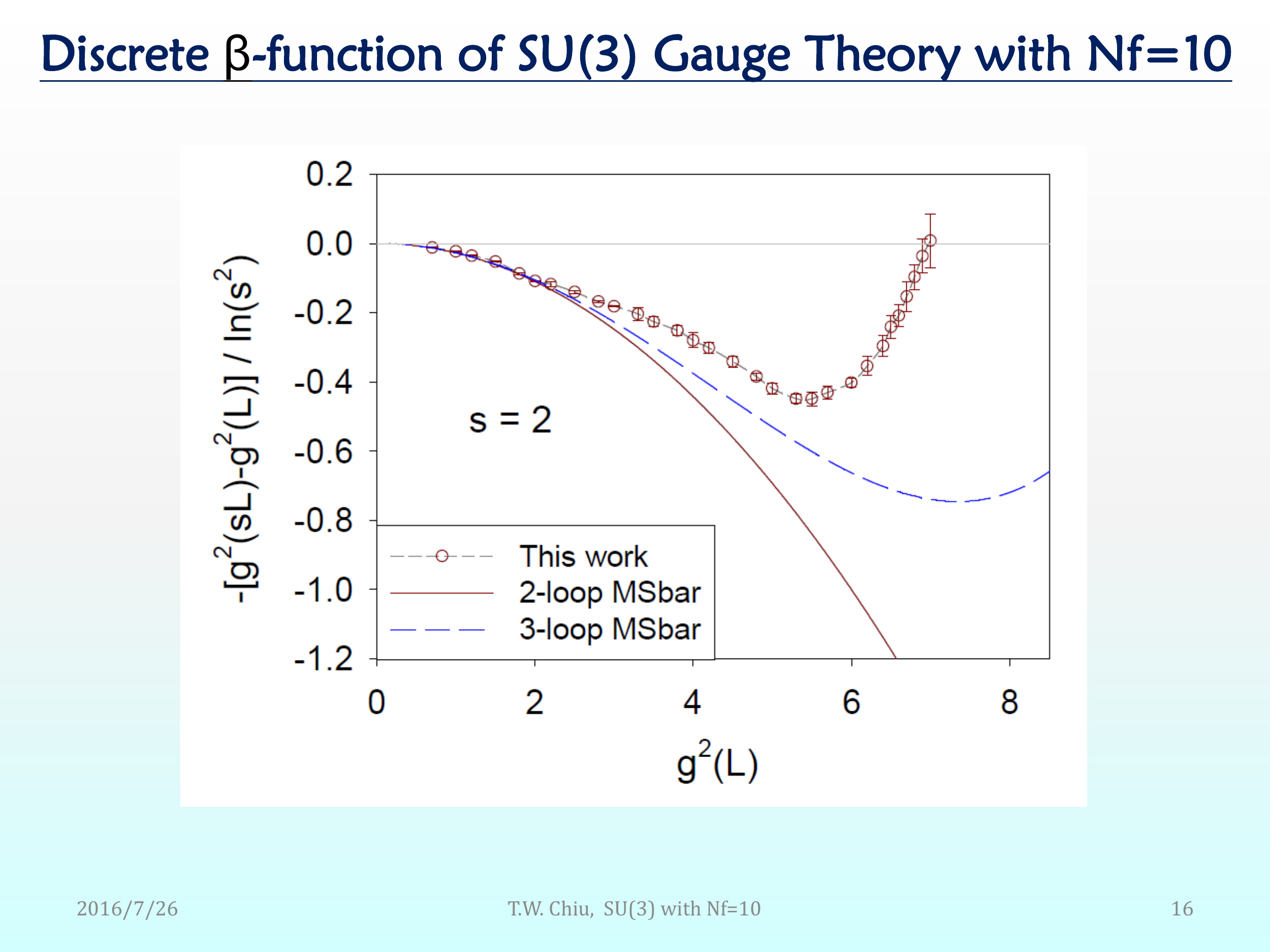}
     \caption{Continuum discrete $\beta$-function for the SU(3) $n_f=10$ model from \cite{chiu:2016}. }
     \label{fig:chiu}
     \end{figure}
For the $n_f=10$ model, an update of \cite{Chiu:2016uui} was presented at this conference \cite{chiu:2016} which includes a larger $L/a=32$ lattice volume. The study measure the $\beta$-function of the model in the finite volume gradient flow coupling scheme, similar to the $n_f=12$ studies above, but it uses an optimal domain-wall fermion action \cite{Chiu:2015sea}. The use of this particular action with good chiral properties seems to allow the use of rather small lattices in a leading order continuum extrapolation with only linear $(a/L)^2$ terms. Using then four steps, the continuum extrapolation is rather well constrained as shown in \fig{fig:chiu}. T
he result clearly points to the existence of an IR fixed point, although a full estimate of systematic errors due to e.g. different choice of discretization of the gradient flow observable, to the choice of interpolation function used for the coupling, or to the choice of continuum extrapolation has not yet been carried out.
If confirmed, this results will also settle the $n_f=12$ case, as it would be very difficult to imagine it is not IR conformal if the $n_f=10$ model is.

 \begin{figure}[b!]
     \center
     \includegraphics[width=.49\textwidth]{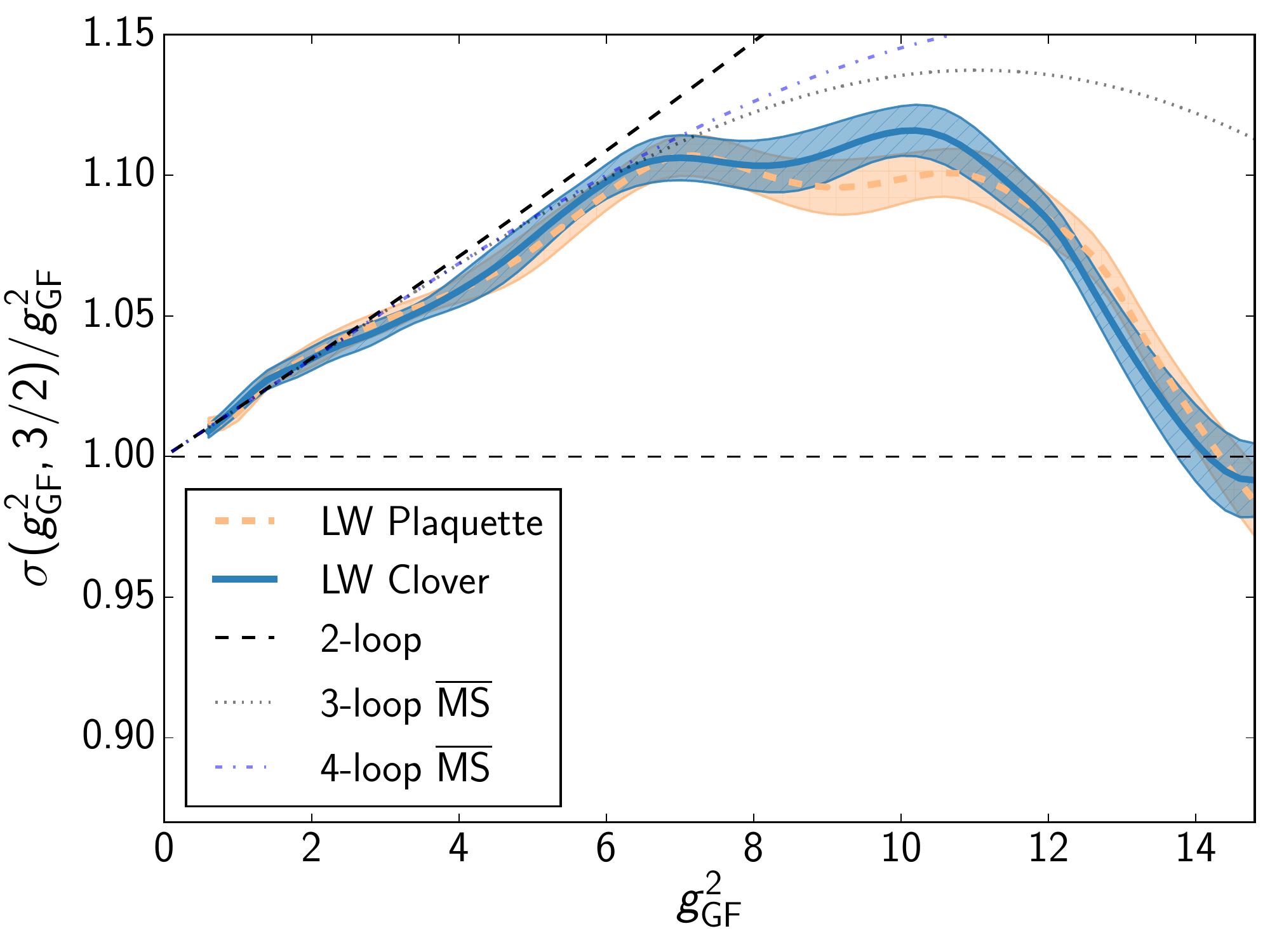}\hfill
     \includegraphics[width=.3\textwidth]{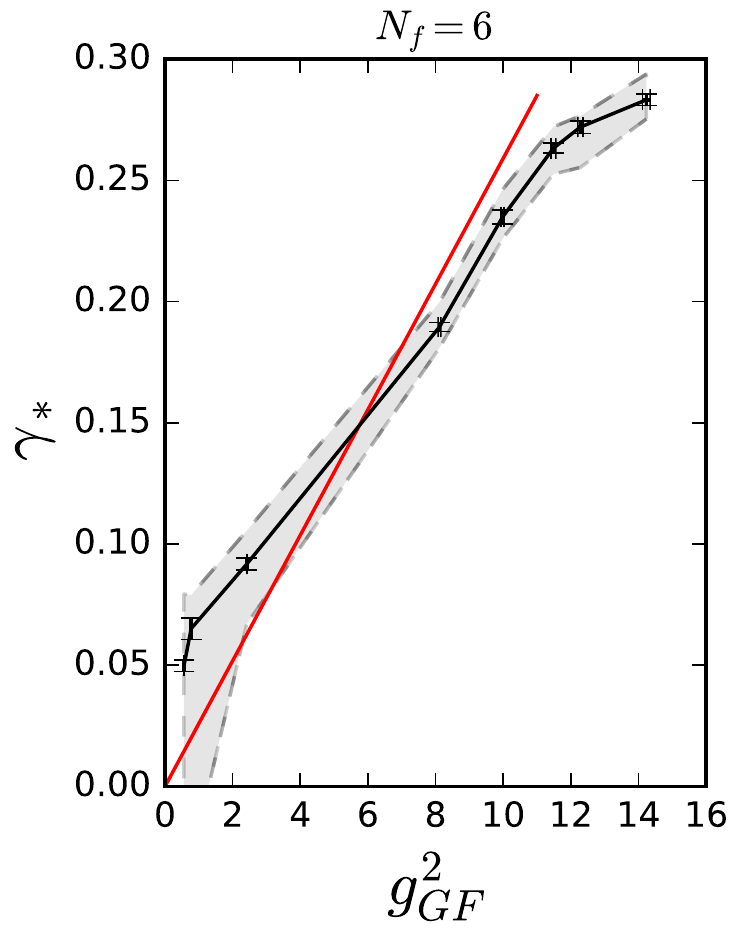}\hspace{2cm}\hfill
     \caption{Left panel: continuum step scaling function for the SU(2) $n_f=6$ from \cite{Leino:2016njf}. Right panel: Mass anomalous dimension from \cite{Suorsa:2016jsf}.}
     \label{fig:hel}
     \end{figure}
\subsection{SU(2) with fundamental fermions}
In the case of two colors, the location of the CW seems more established, as there is agreement among various groups that models with $n_f\geq 8$ lie inside the conformal window whereas models with $n_f\leq 4$ lie outside \cite{Karavirta:2011zg,Ohki:2010sr,Rantaharju:2014ila,Lewis:2011zb,Hietanen:2013fya,Hietanen:2014xca,Arthur:2016dir}. The case $n_f=6$ is more debated and still not settled \cite{Karavirta:2011zg,Bursa:2010xn,Hayakawa:2013yfa,Appelquist:2013pqa}. A new study was presented at this conference \cite{Leino:2016njf,Suorsa:2016jsf} of the non-perturbative $\beta$-function in the SF gradient flow coupling scheme with an HEX-smeared Wilson-clover action which allows to reach rather strong couplings. The authors perform a careful study of the systematic errors involved and conclude in favor of the existence of an IR fixed point at a rather large value of the coupling $g^2\sim 15$, see \fig{fig:hel} for an example of their final result. The same authors also investigate the mass anomalous dimension finding a value at the fixed point of $\gamma\sim 0.3$.
Given the experience with the SU(3) $n_f=12$ case, large volumes will be required to check these results in the vicinity of the observed fixed point and to confirm its existence. In fact this is the region where the raw data shows the most sensitivity to volume, as expected, and can therefore affect the continuum extrapolation in a step-scaling analysis for the extraction of the continuum $\beta$-function.

\subsection{SU(2) with adjoint fermions}
Previous effort focused mainly on the model with $n_f=2$ adjoint Dirac fermions \cite{Catterall:2007yx,Hietanen:2008mr,DelDebbio:2008zf,DelDebbio:2009fd,Catterall:2009sb,DelDebbio:2010hx,DelDebbio:2010hu,Bursa:2011ru,DelDebbio:2015byq}.
The model was found to be IR conformal with an anomalous dimension of the mass\footnote{Different methods for the determination of $\gamma$ do not fully agree with each other, hinting to residual systematic errors not fully under control.} $\gamma\sim 0.2-0.4$.
A previous study of $n_f=1$ Dirac fermion\footnote{The global symmetry in this case is SU(2) which is too small for a model for dynamical EW symmetry breaking.} \cite{Athenodorou:2014eua} presented evidence hinting at the model being inside the conformal window with a rather large mass anomalous dimension $\gamma\sim 1$.
At this lattice conference this study was extended to cover the full range of allowable number of adjoint fermions $n_f=1/2, 1, 3/2, 2$ \cite{bergner:2016}, where half-integer numbers of Dirac fermions represent odd numbers of Weyl fermions.
The case of $n_f=1/2$ massless fermion corresponds to ${\cal N}=1$ supersymmetric Yang-Mills theory.
In particular, the authors of \cite{bergner:2016} presents new results for the $n_f=2, 3/2$ models.
The measured observables include the masses of triplet and singlet mesons, the $0^{++}$ glueball mass, the mass of the lightest composite spin-1/2 fermion-gluon state, and the mass anomalous dimensions at two different values of the lattice spacing.
By studying ratios of masses for different states as a function of the bare quark mass in the chiral limit, evidence was presented that the $n_f=2, 3/2$ models lie inside the conformal window.
The mass anomalous dimension has been determined both from the spectral quantities and from the mode number of the Dirac operator.
The study confirmed the presence of very large finite size effects when approaching the chiral limit, as pointed out in \cite{DelDebbio:2015byq}, which makes numerical simulation close to the chiral limit extremely expensive.
By comparing their results at two different values of the lattice spacing, a residual dependence on the lattice cutoff was observed, in particular for the measured value of the mass anomalous dimension which seems to decrease significantly at smaller lattice spacing.
For example at $n_f=1$ the (preliminary) value of $\gamma$ drops to $~0.75$ on the finer lattice spacing used in this work.
Such residual dependence is observed for all models inside the CW, i.e. $n_f=1,3/2, 2$, and it might be due to residual finite size effects, which drive the system away from the IR conformal fixed point.
More studies are needed to determine the precise value of $\gamma$ for these models.

\begin{figure}[t!]
\centering
\hfill{\includegraphics[height=0.23\textheight]{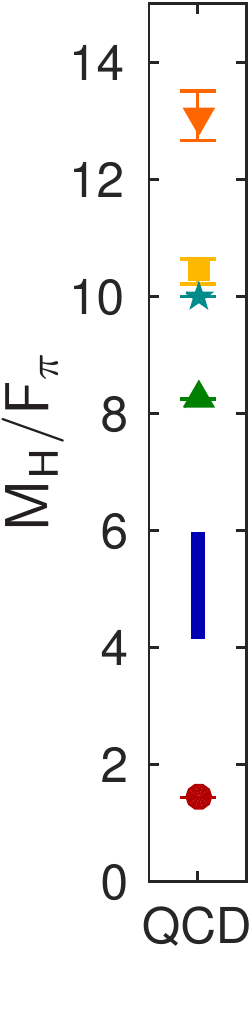}}
\hfill{\includegraphics[height=0.23\textheight]{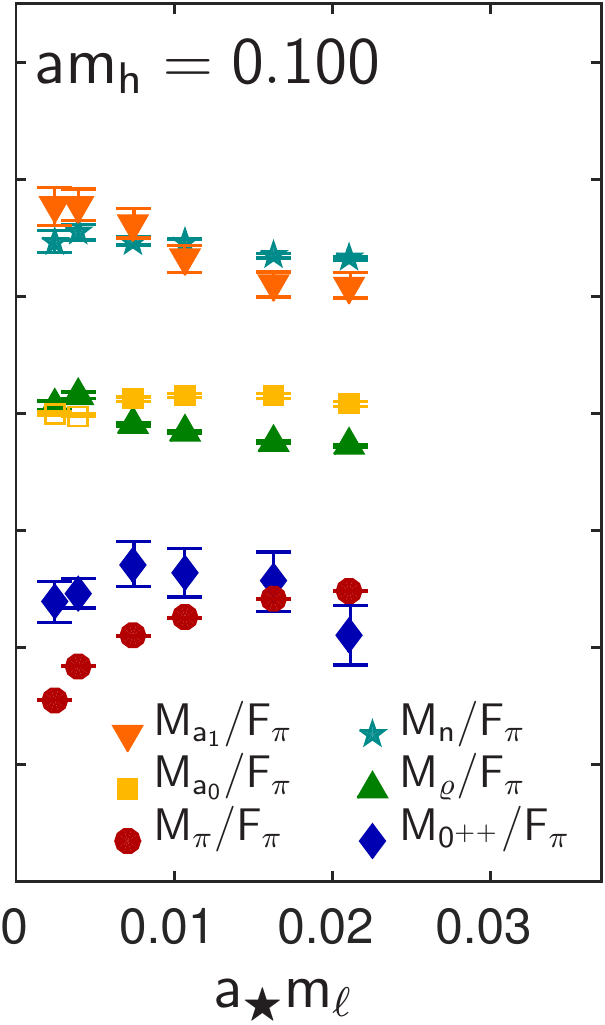}}
\hfill{\includegraphics[height=0.23\textheight]{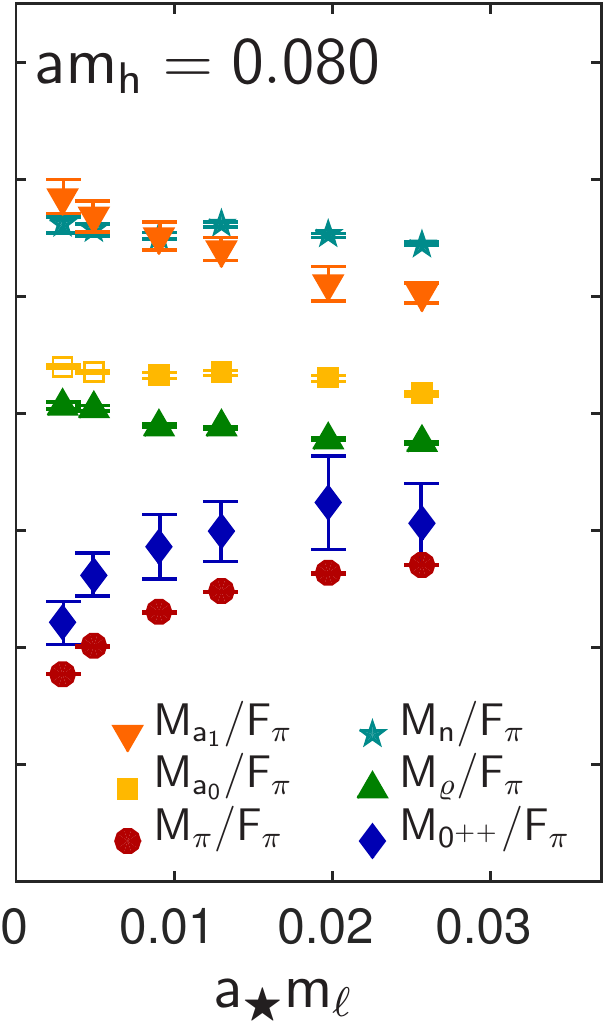}}
\hfill{\includegraphics[height=0.23\textheight]{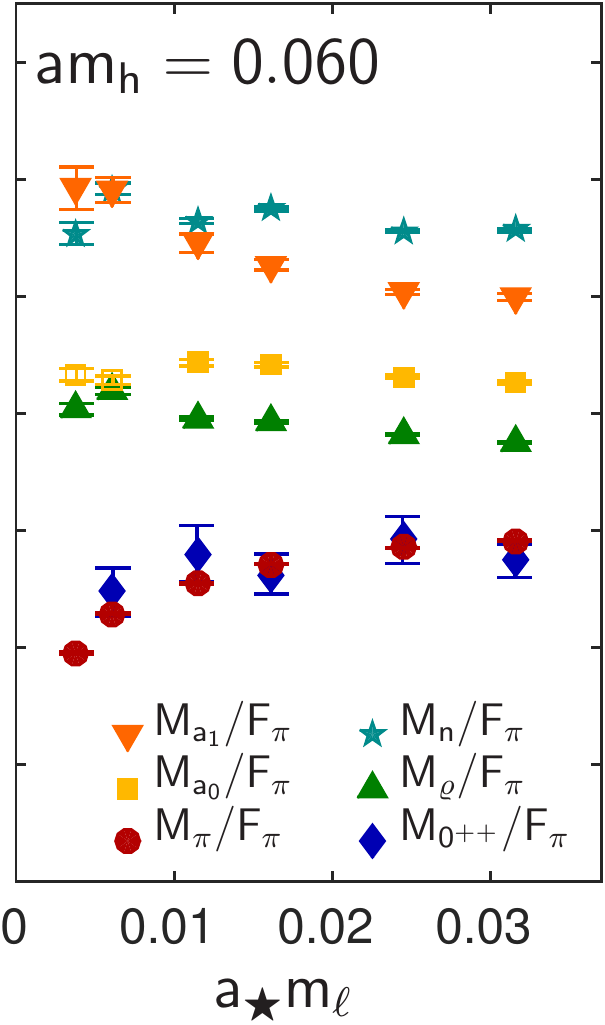}}
\hfill{\includegraphics[height=0.23\textheight]{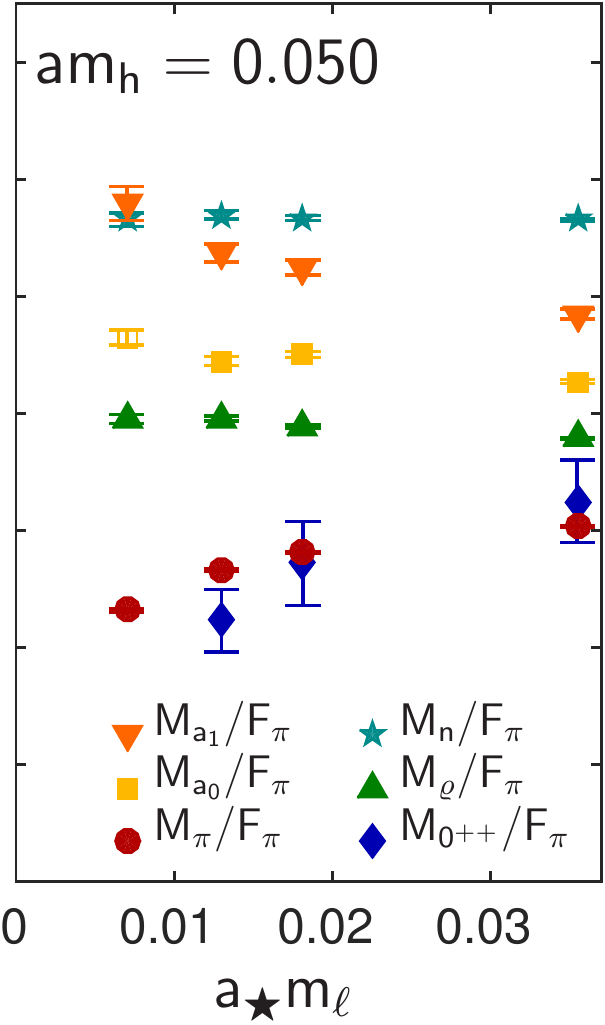}}
\hfill{\includegraphics[height=0.23\textheight]{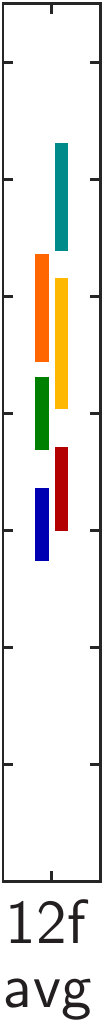}}
\caption{Spectrum of the SU(3) model with $n_f=4l+8h$ fundamental fermions. Each panel corresponds to a different mass value for the heavy fermions; the leftmost panel corresponds to the physical spectrum of QCD while the rightmost panel corresponds to the $n_f=12$ model for a fixed fermion mass. From \cite{Brower:2015owo}.}
\label{fig:rebbi}
\end{figure}

\section{Light composite scalars}\label{sect:scalars}

Several interesting examples of strongly coupled models with light scalars have been presented at this conference.
These are based on SU(3) gauge group with either $n_f=4+8$ or $n_f=8$ fundamental fermions, or with $n_f=2$ fermions in the 2-index symmetric representation (sextet).

\subsection{SU(3) with $n_f=4l+8h$ fundamental fermions}

Based on the idea that walking can be "generated" by continuously deforming a model with IR fixed point, one can introduce a small fermion mass to study the spectrum of the model close to an IR fixed point \cite{DelDebbio:2010hx, DelDebbio:2010hu} in the "walking" regime.
The model considered here is build on the assumption that the SU(3) model with $n_f=12$ fundamental fermion has an IR fixed point (see \sect{sect:su3f}).
One could then introduce a mass $m_h$ for eight of the twelve fermions so that in the limit of large $m_h$ the model is reduced to the $n_f=4$ model which is then similar to ordinary QCD.
For intermediate masses $m_h$, corresponding to a walking regime, the spectrum of the model can be studied.
In actual lattice simulations, it is also necessary to introduce a mass $m_l$ for the four light fermion, so that an extrapolation to zero $m_l$ is required for each $m_h$.
Although not a realistic model of dynamical EW symmetry breaking, this model can be considered as a prototype for models of walking TC and for models of pNGB composite Higgs \cite{Vecchi:2015fma}, in the latter case based on the coset SU(4)$\times$SU(4)/SU(4)\footnote{As noted in \cite{Vecchi:2015fma}, in this case a simple realistic model which takes into account partial compositeness effectively reduces the coset to SU(4)/Sp(4).}.

Extending their previous work \cite{Brower:2015owo}, new results for the spectrum and scaling properties of this model were presented at this conference \cite{Hasenfratz:2016uar, rebbi:2016}.
The spectrum for the light-light mesons as a function of the heavy $m_h$ and light fermion mass $m_l$ is shown in \fig{fig:rebbi}, where meson masses are normalized by the value of the pseudoscalar decay constant $F_\pi(m_l,m_h)$.
In the chiral limit $m_l\to 0$ the ratios $M_H/F_\pi$ seem to depend only weakly on the value of $m_h$, with the exception of the scalar $0^{++}$ state which seems to become much lighter as $m_h$ is reduced, i.e. when entering the "walking regime" of the model.
In fact for $a m_h\leq 0.06$ the scalar state becomes degenerate, within errors, with would be NGBs of the model, which is a common feature observed in numerical simulations of models with light scalar states.
This feature makes it difficult to extrapolate the numerical data for the NBGs and scalar sector to the chiral limit, as the ordinary chiral perturbation theory is not applicable in the regime where numerical simulations can be performed, and one should consider how to include light scalar states in the effective description of the model (see also \sect{sect:nf8} below).
It is therefore still unclear precisely how light such scalar states are.
The light-heavy and heavy-heavy meson states were also investigated in \cite{Hasenfratz:2016uar, rebbi:2016}.
Assuming its existence, close enough to the IR fixed point for the $n_f=12$ model at $m_l=m_h=0$, hyperscaling relations hold for the masses of hadrons and decay constants, and their ratios.
In \cite{Hasenfratz:2016uar, rebbi:2016} evidence is provided for  hyperscaling, which supports the hypothesis of an IR fixed point in the $n_f=12$ system.

\begin{figure}[t!]
     \center
     \includegraphics[width=\textwidth]{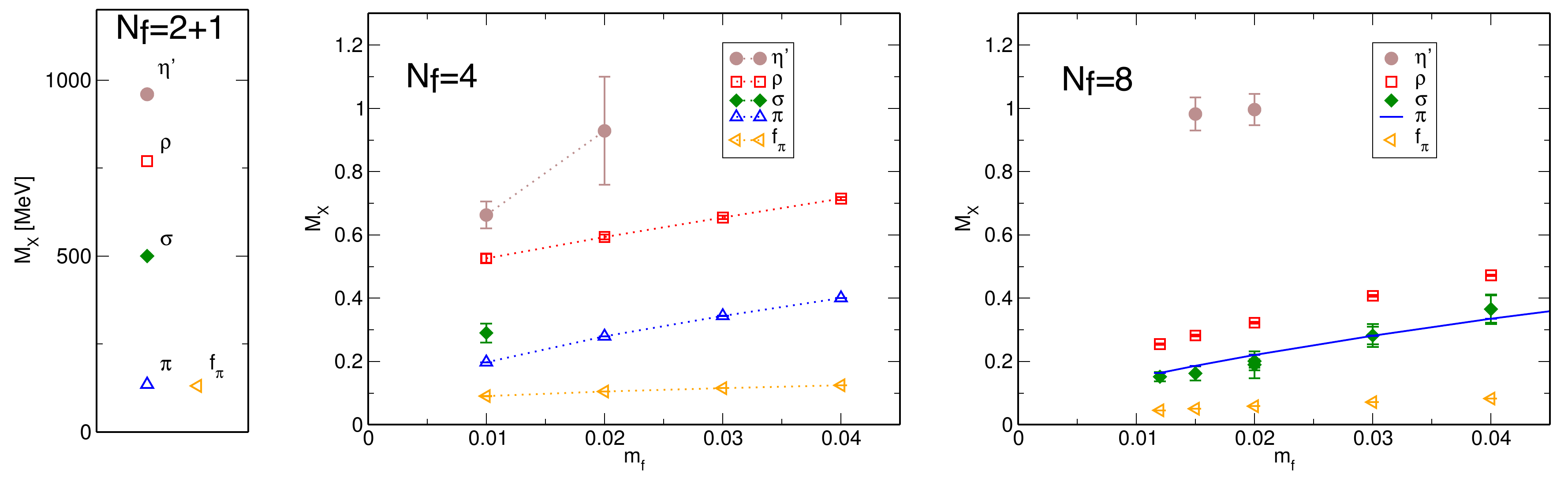}
     \caption{Spectrum of the SU(3) model with $n_f=4,8$ fundamental fermions, as compared to QCD. From the LatKMI collaboration \cite{aoki:2016}. }
     \label{fig:aoki}
     \end{figure}
\begin{figure}[b!]
     \center
     \includegraphics[width=.45\textwidth]{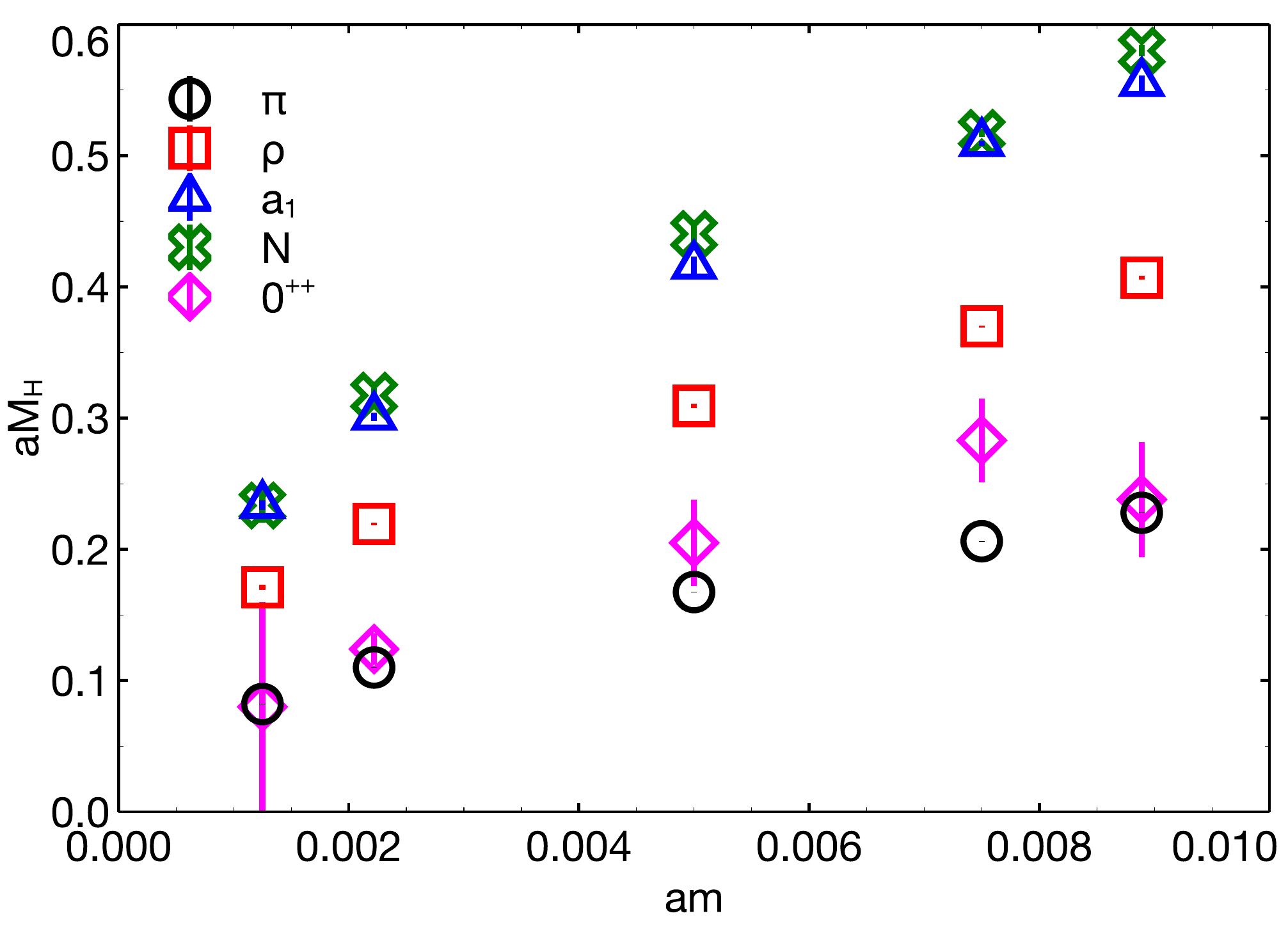}\hfill
     \includegraphics[width=.45\textwidth]{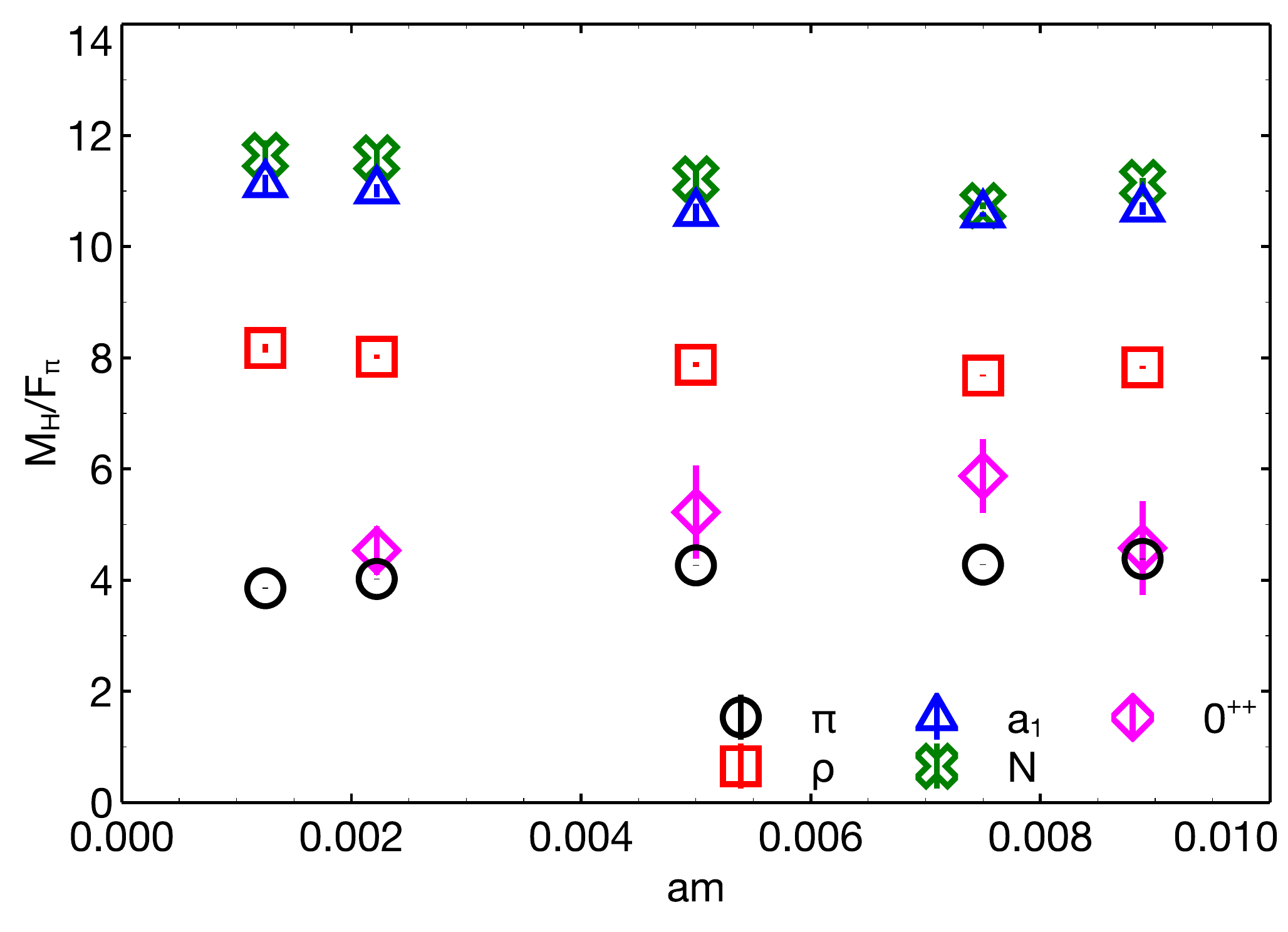}
     \caption{Left panel: Spectrum of the SU(3) model with $n_f=8$ fundamental fermions. Right panel: Same as left panel but in units of $F_\pi$. From the LCD collaboration \cite{fleming:2016}. }
     \label{fig:fleming}
     \end{figure}
\subsection{SU(3) with $n_f=8$ fundamental fermions}\label{sect:nf8}
The results from the previous section point to the possibility of light scalars for walking models which lie just below the onset of the CW.
As a direct test of this hypothesis, different groups have been investigating the spectrum of the SU(3) model with $n_f=8$ fundamental fermions \cite{Aoki:2013xza, Appelquist:2014zsa, Aoki:2014oha,Appelquist:2016viq}.
At this conference both the LatKMI and the LSD collaboration have presented new results for the spectrum of this model.

The LatKMI collaboration presented new results for spectrum of the $n_f=8$ and $n_f=4$ models \cite{aoki:2016, Aoki:2016wnc}, shown in \fig{fig:aoki}, in particular for the flavor singlet mesons $\sigma$ and $\eta'$.
In the $n_f=8$ model, the $\sigma$ meson appears to be degenerate with the pion over the whole range of fermion masses explored, while for the $n_f=4$ case it remains heavier than pion at the most chiral point investigated.
This hints at the possibility of a much lighter $\sigma$ resonance than in QCD.
However if one considers the ratio $m_\sigma/m_\rho$, or the ratio $m_\sigma/F_\pi$, at the most chiral point in \fig{fig:aoki}, this is unchanged between the $n_f=4$ and $8$ models, and it is also similar to the QCD case.
At face value, this constrasts our previous indication and it would indicate that the $\sigma$ resonance does not become lighter respect to the strong scale at the onset of the conformal window.

Similar results were presented by the LSD collaboration \cite{fleming:2016,gasbarro:2016} at a smaller value of the pion mass.
The mass spectrum of the model, as obtained from the LSD collaboration, is shown in the left panel of \fig{fig:fleming}, while in the right panel shows the ratios $m_H/F_\pi$.
Very large volumes up to $64^3\times 128$ were required in order to keep systematic errors under control.
In agreement with the results from the LatKMI collaboration, the mass of the $\sigma$ resonance is found to be degenerate with the would be NG boson of the model and the ratios $m_H/F_\pi$ are very similar to the QCD values, indicating only a very weak dependence of the ratios on the number of flavors $n_f$.

From the right panel of \fig{fig:fleming} one should notice that both the would be NG boson and the light scalar resonance mass in units of $F_\pi(m_q)$ show only a very weak dependence on the fermion mass $m_q$ in the region explored.
If the $n_f=8$ is not inside the conformal window, then a sharp decrease of $m_\pi/F_\pi$ is expected close to the chiral limit, while the $\sigma$ resonance should remain massive.

It is therefore crucial to be able to extrapolate the current results to the chiral limit to establish if the scalar $\sigma$ resonance becomes much lighter for walking models.
The difficulty stems from the fact that the usual chiral perturbation theory cannot be trusted in the presence of a scalar state as light as the pion.
One attempt to develop a more appropriate effective description which takes into account the presence of a light scalar states was prensented at this conference \cite{Golterman:2016eth} based on previous results by the same authors in \cite{Golterman:2016lsd}.
In this new effective model, the authors consider the case of dilaton scalar state and develop a systematic expansion in the three small parameters $m_q$, 1/N, and $\hat{n}_f-\hat{n}_f^*$, assumming that scale invariance is recovered in the limit $m_q\to0$, large-N Veneziano limit, and $\hat{n}_f/\hat{n}_f^*\to1^-$, where $\hat{n}_f=n_f/N$ and $\hat{n}_f^*$ is the critical number of flavors, in the Veneziano large-N limit, for which the conformal window opens.
Results for $m_\pi$, $m_\sigma$ and the fermion condensate were obtained at leading order, that predict a distinctive behavior as a function of the fermion mass which can in principle be compared to lattice data and it could therefore be a useful analytic tool to test the dilaton-Higgs scenario in walking TC models.

However one must bear in mind that the validity of the effective theory is limited to small values of the three expansion paramters.
Since a naive estimate indicates a value of $\hat{n}_f^*\sim 4$, in the present case of $n_f=8$ this corresponds to $\hat{n}_f^*-\hat{n}_f\simeq 1.3$, which might be too large for the expansion to hold.
Moreover, given the use of the Veneziano limit, the effective model cannot be used without modifications for the case of two-index representations.

Other approaches to an effective description of the $\pi$-$\sigma$ system also exist such as \cite{Soto:2011ap, Matsuzaki:2013eva, Hansen:2016fri}.

\subsection{SU(3) with $n_f=2$ sextet fermions}

\begin{figure}[t!]
     \center
     \includegraphics[width=.8\textwidth, height=5.8cm]{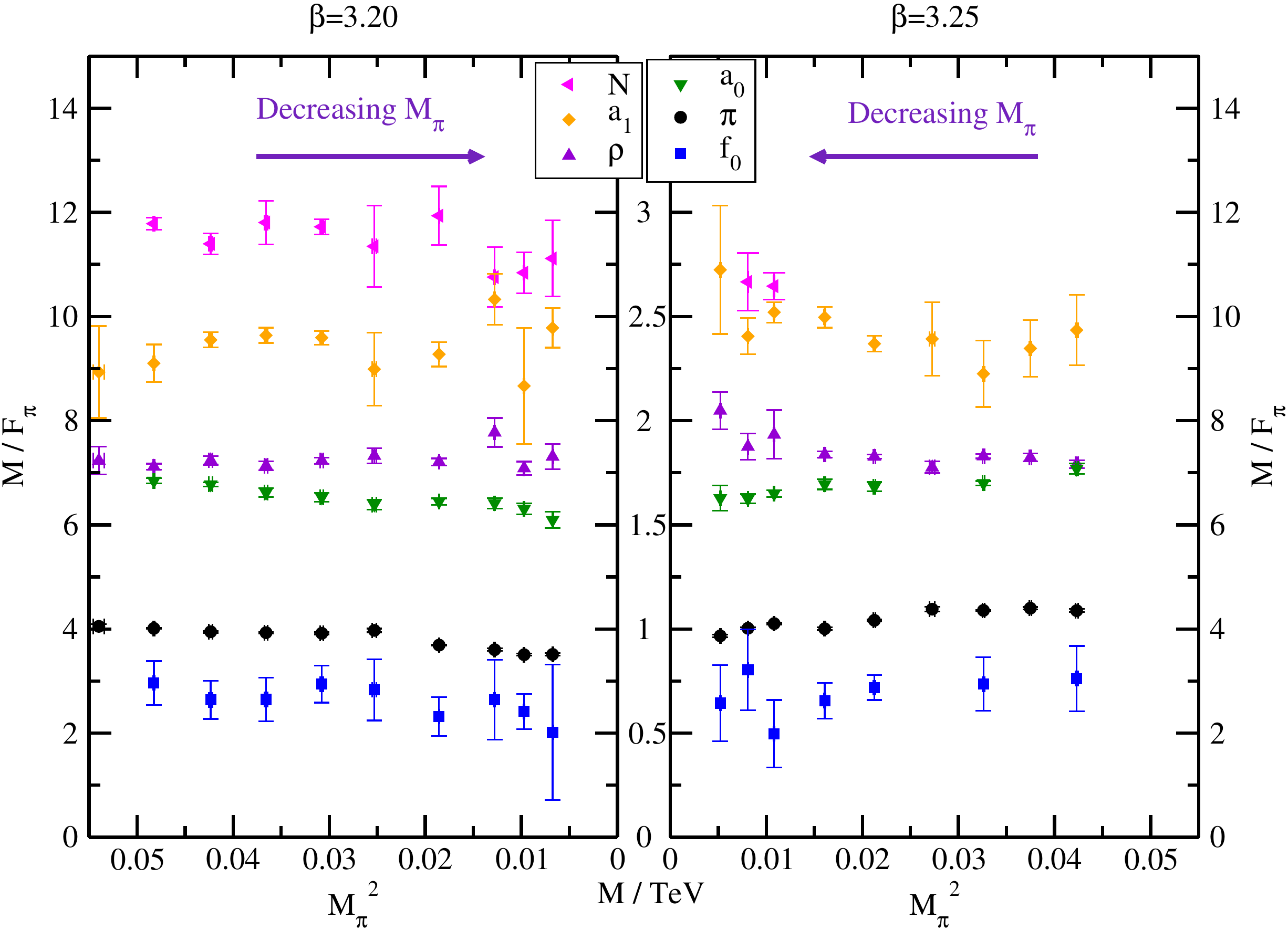}
     \caption{Spectrum of the SU(3) sextet model with $n_f=2$ two-index symmetric fermions in units of $F_\pi$.  From \cite{Fodor:2016pls}. }
     \label{fig:kuti}
     \end{figure}

Another very interesting walking TC model is based on SU(3) with $n_f=2$ fermions in the two index symmetric representation (the sextet representation).
The model has the three NGBs, i.e. the minimal number required for a TC model, but from the higher dimensional representation of the fermions, one expects the model to be walking \cite{Dietrich:2005jn}, with possible light scalar states.
The spectrum of this sextet model has been studied in detail on the lattice with staggered fermions \cite{Fodor:2012ty, Fodor:2014pqa, Fodor:2016wal, Fodor:2016pls} and more recently an investigation with Wilson fermions also began \cite{Drach:2015sua}.
Updates for both lattice formulations have been presented at this conference \cite{kuti:2016,wong:2016,Hansen:2016sxp}.

I report in \fig{fig:kuti} the spectrum of the model as obtained by staggered fermions lattice simulations, at two different lattice spacings \cite{Fodor:2016pls}.
The qualitative features of the spectrum are similar to the other walking models described above.
The spectrum features a light $\sigma$ resonance over the entire range of light quark masses explored, in fact lighter that the would be NGB, which should eventually become massless in the chiral limit for a chirally broken model.

There is still some controversy about the sextet model being inside or outside the conformal window.
From an eye inspection of \fig{fig:kuti}, one might be tempted to conclude that the model is in fact IR conformal and inside the conformal window, however a more detailed analysis has led the authors of \cite{Fodor:2016pls} to conclude that this is not the case as the data do not follow the hyperscaling predictions.
On the other hand the data seems to fits well with the prediction of rooted staggered chiral perturbation theory, even if the use of such effective model is not justified,  given the presence of the light scalar state.
As probing the model at even lighter masses in the $p$-regime would be prohibitively expensive, the authors of \cite{Fodor:2016pls} are moving to use more sophisticated analysis methods involving the cross-over regime from the $p$ to the $\epsilon$-regime, the use of random matrix theory and the use of effective models which take into account the light scalar particle in the spectrum.
\begin{figure}[b!]
     \center
     \includegraphics[width=.49\textwidth]{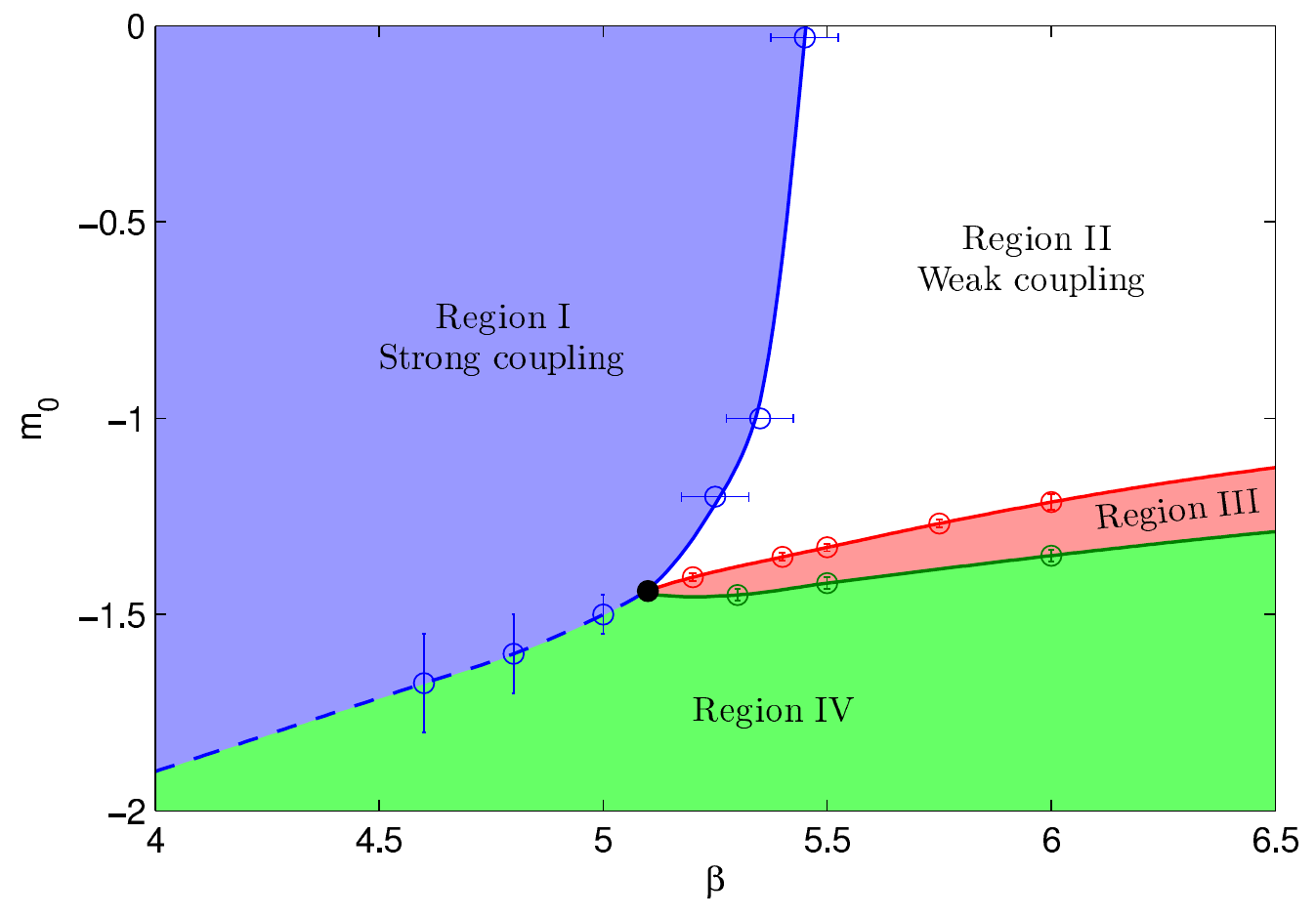}\hfill
     \includegraphics[width=.478\textwidth]{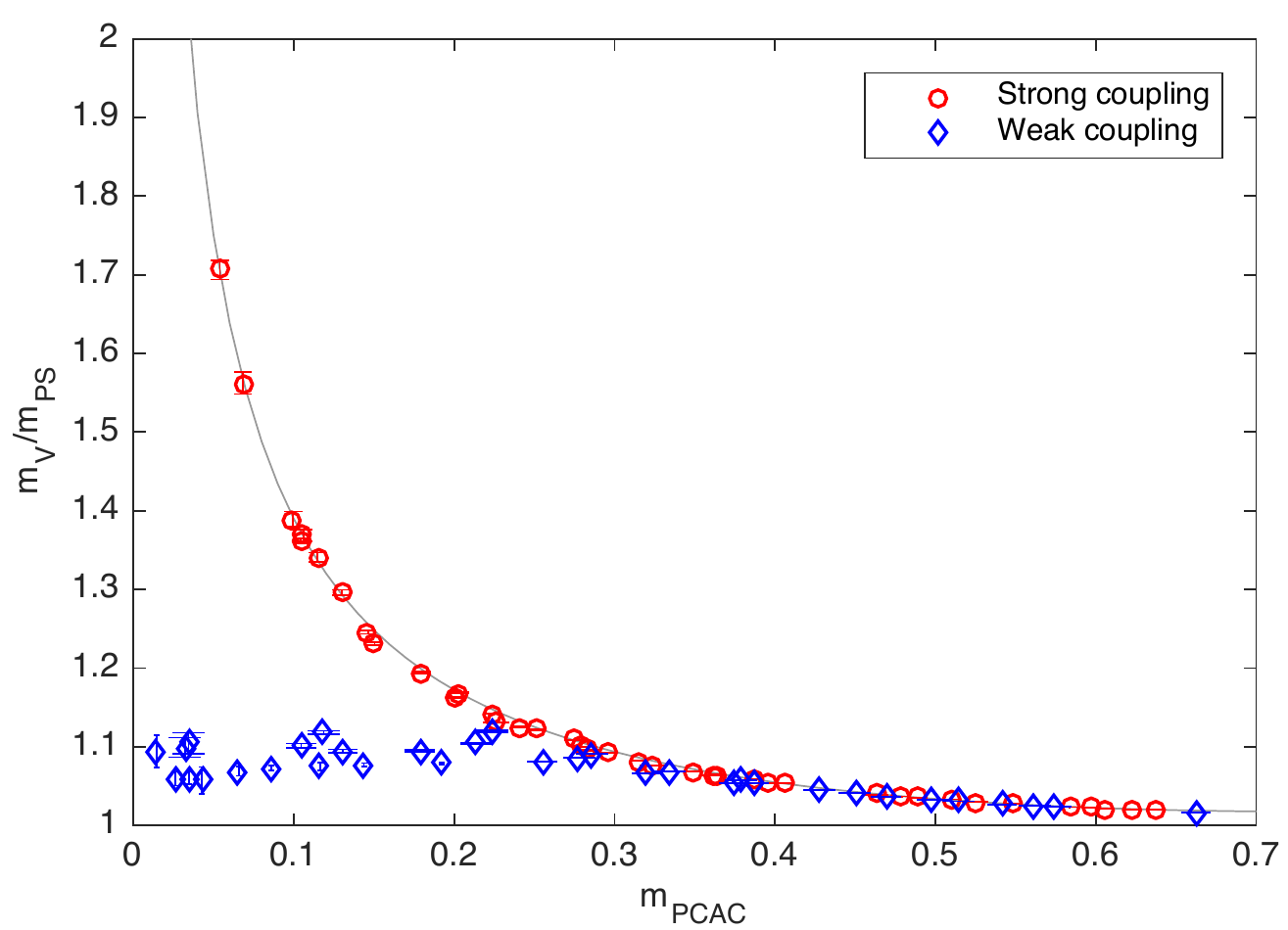}
     \caption{Left panel: Phase structure of the lattice SU(3) sextet model with Wilson fermions. Right panel: The ratio $m_\rho/m_\pi$, for different lattice spacings in the weak coupling and strong coupling phases. From \cite{Hansen:2016sxp}. }
     \label{fig:hansen}
     \end{figure}

The spectrum of the sextet model is also being investigated with Wilson fermions.
Similarly to the staggered fermion case,  the mass spectrum can be fitted to a prediction from Wilson chiral perturbation theory, although in this case one cannot also exclude the possibility of hyperscaling from an IR fixed point \cite{Drach:2015sua}.
There is however a striking difference in the between the spectra from staggered and Wilson fermions: in the latter case, the vector resonance never appears to become much heavier that the pion in the weak coupling phase.

To better understand the behavior of the model, a new study of the full phase structure of the lattice model with Wilson fermions was presented at this conference \cite{Hansen:2016sxp}, see \fig{fig:hansen}.
A phase at strong coupling was identified, separated from the weak coupling phase by a crossover.
At strong coupling there is a first order transition as the fermion mass is reduced, which becomes a continuous transition in the weak coupling phase corresponding to the line of vanishing PCAC mass.
The behavior of several quantities, which include the mass spectrum and the scale-setting observables $w_0$ and $t_0$, was studied both in the weak coupling and strong coupling phase.
The results show a sharp change in the qualitative behavior of the measured quantities, see e.g. the right panel of \fig{fig:hansen}.
While at strong coupling, the observations are compatible with a chirally broken model as expected, data in the weak coupling phase do not show any clear indications of spontaneous chiral symmetry breaking.

The question of the infrared conformality of this model will require the use of more data at
weak coupling on the spectrum of the model, at several lattice spacings, to show consistently the
presence or not of a critical behavior in the chiral limit.

\section{pNGB Higgs models}\label{sect:pngbhiggs}

Several interesting pNGB Higgs models have been considered at this conference, which cover the three minimal cosets discussed in \sect{sect:ngh}.
The case of SU(4)$\times$SU(4)/SU(4) can be realized with fundamental fermions of SU(3), and it has been discussed above.
Here we discuss the contributions related to the other two cases.

\subsection{SU(4) with sextet and fundamental fermions}
\begin{figure}[t!]
     \center
     \includegraphics[width=.51\textwidth]{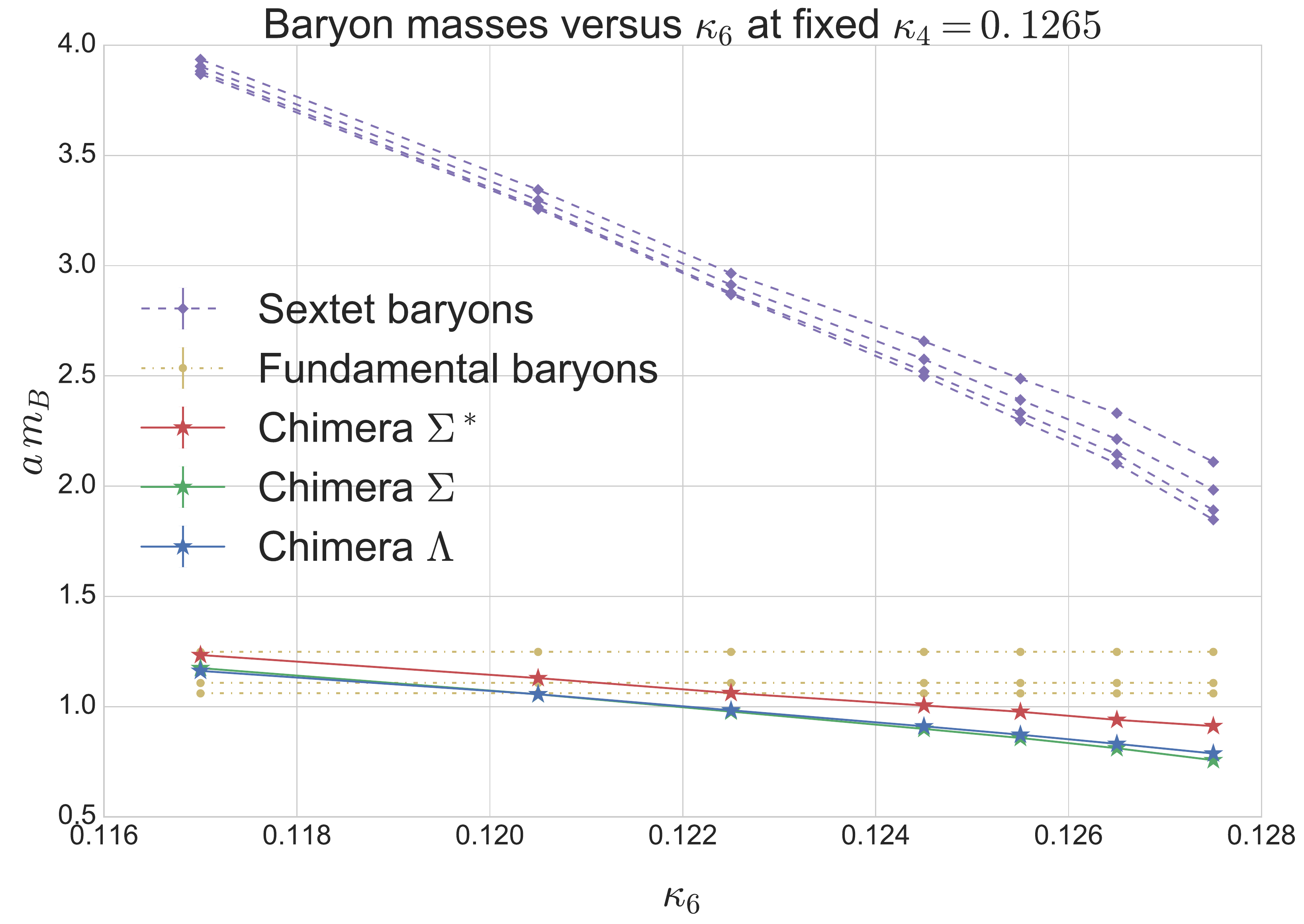}
     \caption{Spectrum of the SU(4) model with $n_f=2$ fundamental fermions and quenched sextet fermions. ``Chimera'' baryons are composite of quarks in the two representations. From \cite{DeGrand:2016mxr}. }
     \label{fig:jay}
     \end{figure}

The case of the SU(5)/SO(5) coset can be realized with five Majorana fermions in two-index antisymmetric (sextet) represetation of SU(4).
This coset has also been suggested as the base for a model of partial compositeness with three additional Dirac fundamental fermions \cite{Ferretti:2016upr}.
The odd number of fermions makes this model harder to study via lattice simulations.
As a first step, the case of SU(4) with four Majorana in the two-index antisymmetric represetation and the case of SU(4) with two fundamental fermions plus quenched two-index antisymmetric fermions were studied \cite{DeGrand:2016mxr,DeGrand:2016het}.
For the latter case, the spectrum of the model was presented in \cite{DeGrand:2016mxr}, which is shown in \fig{fig:jay}.
It is possible to have baryons which are composite of fermions in both representations, and the quenched study shows that these can become lighter than baryons made of fermion in one representation only.
This could be interesting for model building of light top-partners.
However this result in the quenched approximation is not conclusive and the full spectrum of the model should be considered with dynamical fermions.

In \cite{DeGrand:2016het} the radiative contributions from electroweak gauge bosons to the composite Higgs potential of the SU(4) model with four Majorana fermions in the sextet representation were considered.
This calculation is similar to the electromagnetic contribution to the masses of the pions in QCD.
Electroweak gauge boson generate a potential for the composite Higgs $h$ of the form: $V(h)=C_{LR} (3g^2+g'^2)(h/F_\pi)^2+\mathcal{O}(h^4)$ where the positive constant $C_{LR}$ can be computed from a vacuum polarization function $\Pi_{LR}$.
A test of the feasibility of the measure of $C_{LR}$ on the lattice was presented in \cite{DeGrand:2016het} with two different approaches found to be in good agreement.
In the future this approach will be extended to the full model with fermions in two different representations.

One should remember that potential generated by EW gauge bosons will not generate a \textit{vev} for the higgs field and will not break electroweak symmetry, as $C_{LR}>0$.
For this to occur, radiative corrections from the interactions with SM fermions, in particular the top quark, should be included.
As discussed above in \sect{sect:ngh}, for the composite pNGB Higgs scenario to be viable the Higgs \textit{vev} should be small compared to the strong scale $F_\pi$, implying a small alignement angle $\theta$, which requires some subtle cancellation in the effective Higgs potential between the electroweak contibutions and the SM fermion sector.

\subsection{SU(2) with $n_f=2$ fundamental fermions}
This model is the minimal realization of a composite pNGB model, requiring only two fundamental fermions of SU(2).
The model can be used as the building block for a fully realistic composite Higgs model \cite{Cacciapaglia:2014uja} compatible with the experimental constraints if, roughly, $\theta<0.2$ (see \cite{Arbey:2015exa} for details).
The model can also be used to build a model of partial compositeness for all SM fermions \cite{Sannino:2016sfx}, which features hyper-colored scalars and it is free from Landau poles up to the Plank scale.
In a different context, the same model has also been considered as a model of composite dark matter.

\begin{figure}[t!]
     \center
     \includegraphics[width=.49\textwidth]{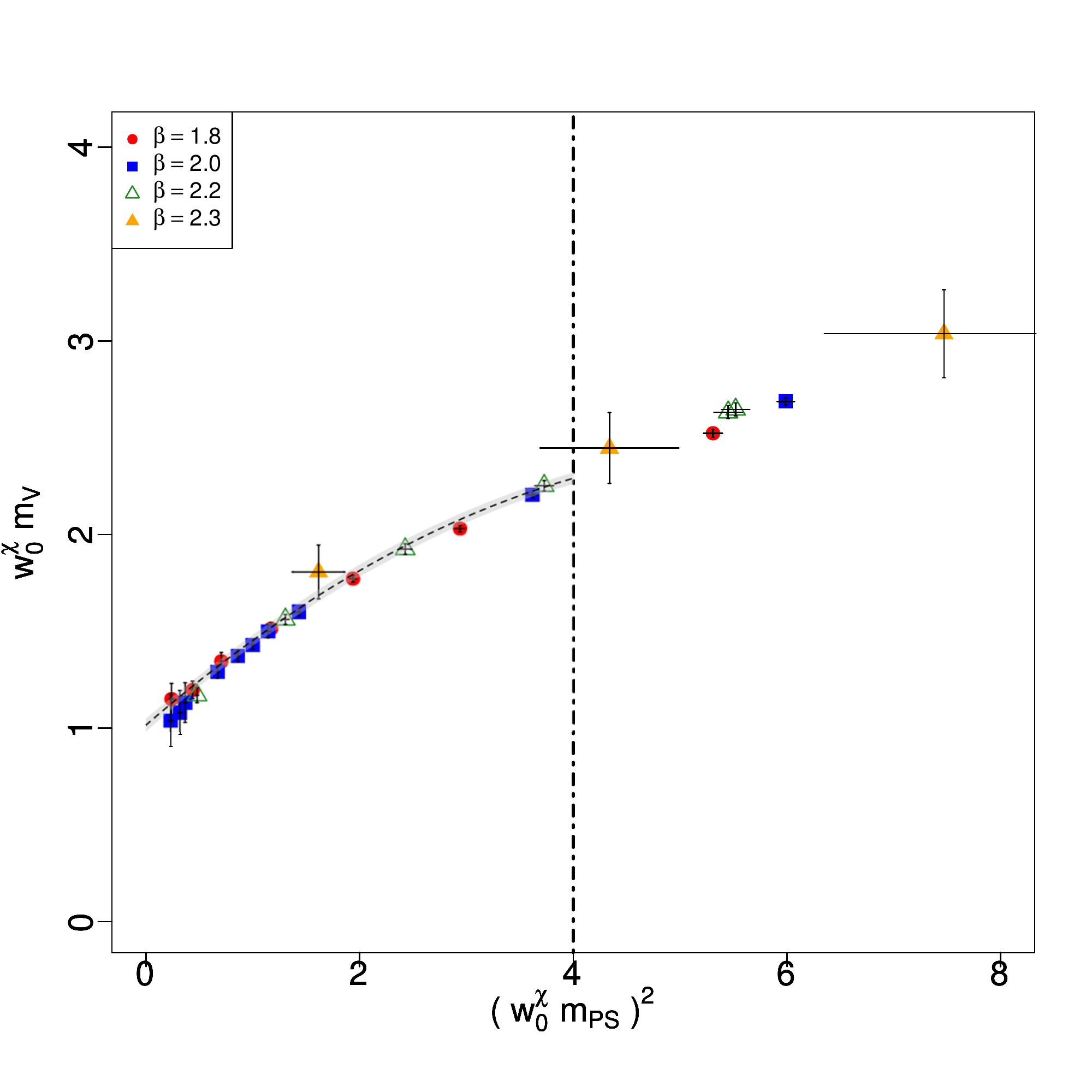}\hfill
     \includegraphics[width=.49\textwidth]{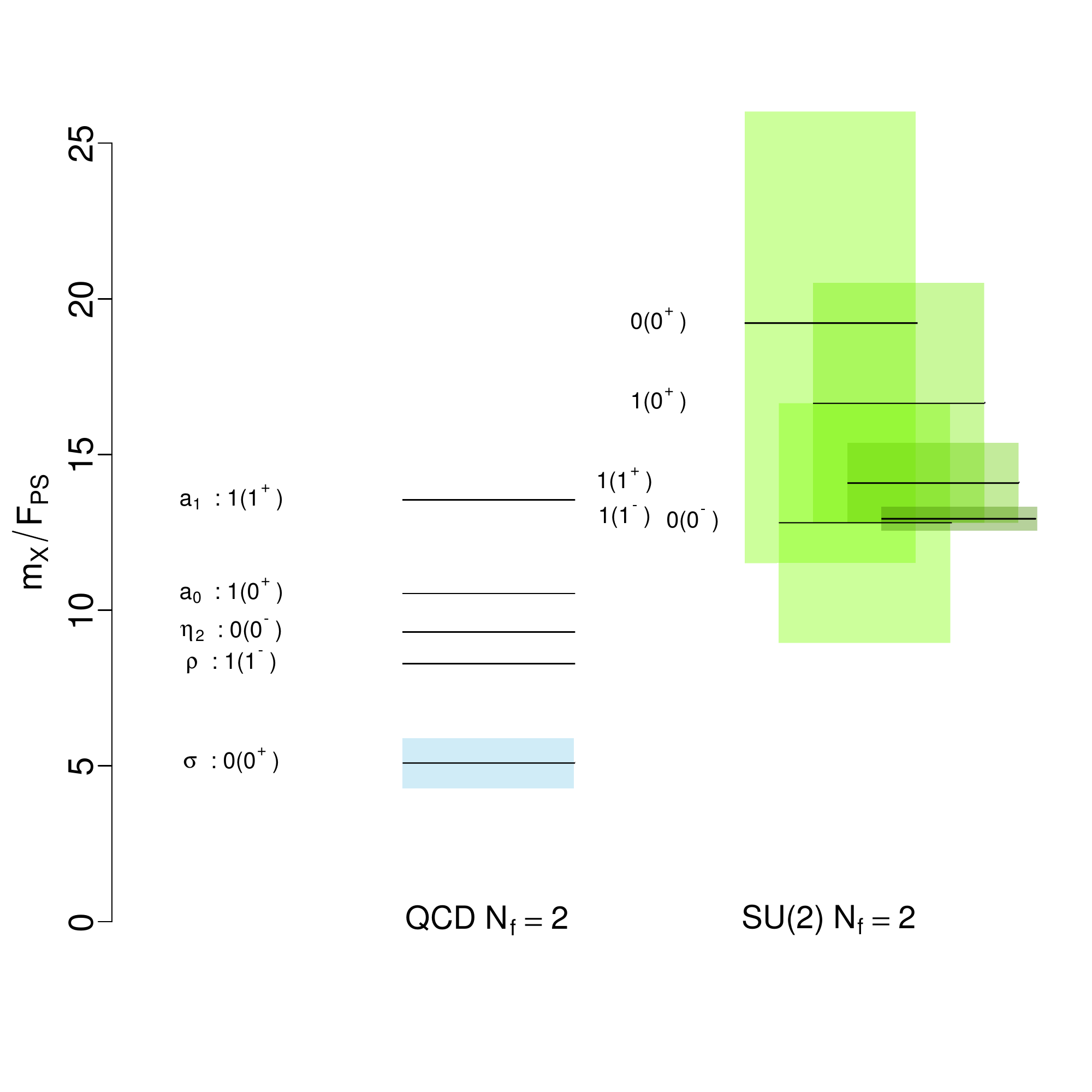}
     \caption{Left panel: Example of combined chiral and continuum extrapolation for the $\rho$ meson. Right panel: Summary plot comparing the spectrum of the SU(2) pNGB model to QCD. From \cite{Arthur:2016dir, Arthur:2016ozw}. }
     \label{fig:drach}
     \end{figure}
Lattice simulations of this model are straightforward, requiring only two colors, two Dirac fermions and no high dimensional representations.
The challenge is then to perform all the necessary extrapolations, i.e infinite volume, zero quark mass and continuum limit for the spectrum of the model, as this has not been done for any other BSM model studied so far.

In \cite{Drach:2016,Janowski:2016hsg} an update for the spectrum of the model was presented, based on \cite{Arthur:2016dir, Arthur:2016ozw}.
Lattice simulations were performed at four different lattice spacings, for a number of quark masses at each lattice spacing, while keeping large enough volumes to reduce finite-volume effects as much as possible.
The scale was set by using the $w_0$ observable \cite{Borsanyi:2012zs} and the RI-MOM scheme \cite{Martinelli:1994ty} was used to measure the required non-perturbative renomalization costants.

Continuum extrapolated results were obtained for $F_\pi$, the fermion condensate, and the lightest spin one and zero resonances, analogue to the QCD $\rho$, $a_1$, $\sigma$, $\eta'$, $a_0$ resonances.
A combined chiral and continuum extrapolation was used to extract the physically interested quantities.
The left panel of \fig{fig:drach} shows an example of such extrapolation for the $\rho$ meson.

The final spectrum for the model is summarized in the right panel of \fig{fig:drach}, in units of $F_\pi$ and compared to the QCD spectrum.
Taken at face value, these results indicate a spectrum which is quite different from the QCD one, featuring heavier resonances which are beyond the present LHC constraints, even in the Technicolor limit of $\theta=\pi/2$.
In the pNGB limit, for $\sin\theta<0.2$, these resonances seem beyond the reach of LHC experiments.
These results are still affected by large systematic errors, as shown in \fig{fig:drach}, mainly due to the chiral and continuum extrapolations required to obtain phenomenological predictions.
The accuracy of these results will be increased in the future.

\section{Conclusions}

The lattice community is actively investigating interesting models for BSM physics.
These non-perturbative studies complement the phenomenological approach by providing valuable information on the strongly coupled dynamics.

The numerical studies of many such models present great numerical and theoretical challenges, most crucially stemming from the near conformal nature of the model requiring very large volumes and the presence of light scalars making the chiral extrapolation difficult.
Nonetheless thanks to the continuous numerical effort and the development of new techniques and tools significant progress has been made in the last few years.

Many results are already available for models based on the gauge groups SU(2) or SU(3) with fermions in the fundamental or higher representations.
A great effort is undergoing to determine the precise location of the conformal window for two or three color models, and only a few borderline cases remaining elusive.
Interesting models featuring light composite scalar states are being investigated in detail and many results for the spectrum of these WTC or pNGB Higgs models are already available.

\acknowledgments
I wish to thank organizers of the Lattice 2016 conference for the kind hospitality.
The work of CP is supported by the Danish National Research Foundation under grant number DNRF90 and by a Lundbeck Foundation Fellowship grant.

\end{document}